\documentclass[namedreferences]{solarphysics}

\usepackage[hyperref,optionalrh]{spr-sola-addons} 
\usepackage{graphicx}        
\usepackage{color}           
\usepackage{breakurl}        
\usepackage[dvipsnames]{xcolor}

\newcommand{\aap}{    {\it Astron. Astrophys.}}

\newcommand{\aj}{     {\it Astron. J.}} 
\newcommand{\apj}{    {\it Astrophys. J.}}
\newcommand{\apjl}{   {\it Astrophys. J. Lett.}}

\newcommand{\mnras}{  {\it Mon. Not. Roy. Astron. Soc.}}

\newcommand{\solphys}{{\it Solar Phys.}}

\chardef\us=`\_

\begin{document}

\begin{article}
\begin{opening}

\title{The First Survey of Quiet Sun Features Observed in Hard X-Rays With NuSTAR}

\author[addressref={aff1},corref,email={s.paterson.5@research.gla.ac.uk}]{\inits{S.}\fnm{Sarah}~\lnm{Paterson}\orcid{0000-0003-2147-9586}}
\author[addressref=aff1]{\inits{I. G.}\fnm{Iain G.}~\lnm{Hannah}\orcid{0000-0003-1193-8603}}
\author[addressref=aff2]{\inits{B. W.}\fnm{Brian W.}~\lnm{Grefenstette}\orcid{0000-0002-1984-2932}}
\author[addressref={aff1,aff3}]{\inits{H. S.}\fnm{Hugh S.}~\lnm{Hudson}\orcid{0000-0001-5685-1283}}
\author[addressref={aff3,aff4}]{\inits{S.}\fnm{Säm}~\lnm{Krucker}\orcid{0000-0002-2002-9180}}
\author[addressref=aff5]{\inits{I.}\fnm{Lindsay}~\lnm{Glesener}\orcid{0000-0001-7092-2703}}
\author[addressref=aff6]{\inits{I.}\fnm{Stephen M.}~\lnm{White}\orcid{0000-0002-8574-8629}}
\author[addressref=aff7]{\inits{I.}\fnm{David M.}~\lnm{Smith}\orcid{0000-0002-0542-5759}}

\address[id=aff1]{School of Physics \& Astronomy, University of Glasgow, University Avenue, Glasgow G12 8QQ, UK}
\address[id=aff2]{Cahill Center for Astrophysics, California Institute of Technology, 1216 East California Boulevard, Pasadena, CA 91125, USA}
\address[id=aff3]{Space Sciences Laboratory University of California, Berkeley, CA 94720, USA}
\address[id=aff4]{University of Applied Sciences and Arts Northwestern Switzerland, 5210 Windisch, Switzerland}
\address[id=aff5]{School of Physics \& Astronomy, University of Minnesota Twin Cities, Minneapolis, MN 55455, USA}
\address[id=aff6]{Air Force Research Laboratory, Space Vehicles Directorate, Kirtland AFB, NM 87123, USA}
\address[id=aff7]{Santa Cruz Institute of Particle Physics and Department of Physics, University of California, Santa Cruz, CA 95064, USA}

\runningauthor{Paterson et al.}
\runningtitle{NuSTAR Quiet Sun}

\begin{abstract}
We present the first survey of quiet Sun features observed in hard X-rays (HXRs), using the the Nuclear Spectroscopic Telescope ARray (NuSTAR), a HXR focusing optics telescope. The recent solar minimum combined with NuSTAR's high sensitivity has presented a unique opportunity to perform the first HXR imaging spectroscopy on a range of features in the quiet Sun.  By studying the HXR emission of these features we can detect or constrain the presence of high temperature ($>$5 MK) or non-thermal sources, to help understand how they relate to larger more energetic solar phenomena, and determine their contribution to heating the solar atmosphere. We report on several features observed in the 28 September 2018 NuSTAR full-disk quiet Sun mosaics, the first of the NuSTAR quiet Sun observing campaigns, which mostly include steady features of X-ray bright points and an emerging flux region which later evolved into an active region, as well as a short-lived jet. We find that the features' HXR spectra are well fitted with isothermal models with temperatures ranging between 2.0--3.2 MK. Combining the NuSTAR data with softer X-ray emission from Hinode/XRT and EUV from SDO/AIA we recover the differential emission measures, confirming little significant emission above 4 MK. The NuSTAR HXR spectra allow us to constrain the possible non-thermal emission that would still be consistent with a null HXR detection. We found that for only one of the features (the jet) was there a potential non-thermal upper limit capable of powering the heating observed. However, even here the non-thermal electron distribution had to be very steep (effectively mono-energetic) with a low energy cut-off between 3--4 keV. 



\end{abstract}
\keywords{Corona, Quiet; Heating, Coronal; Jets; Spectrum, X-Ray; X-Ray Bursts, Hard}
\end{opening}

\section{Introduction}
     \label{S-Introduction} 
The study of the hard X-ray (HXR) emission from the quiet Sun could provide insight into the source of the sustained high temperature of the solar corona, termed the coronal heating problem. It was suggested by \cite{parker} that the source of this heating could be a large number of small-scale energy release events, taking place all through the solar cycle. If such events were weaker versions of flares, they would be expected to produce a HXR signature. 
\par
One phenomenon that has been linked to coronal heating is coronal bright points (CBPs), small-scale loop structures located in the lower corona, which are observed in EUV and soft X-rays (SXRs) \citep[e.g.][]{madjarska}. These features are observed throughout the solar cycle, including solar minimum, when the Sun is quiet in the absence of active regions and large flares. Bright points have been studied extensively in EUV and SXRs. In EUV, these features typically have lifetimes of $<$ 20 hours \citep{alipour,zhang}, whereas they have been found to be shorter-lived in SXRs, with lifetimes of $\sim$ 12 hours \citep{harvey}. These features sometimes have associated transient phenomena, such as ``microflares'' \citep{golub,shimojo}, and small-scale eruptions \citep{mou}.
\par
Previous studies have used EUV and SXR observations to investigate the temperatures of bright points. \cite{kariyappa} used SXR data from the \emph{Hinode} X-ray Telescope (Hinode/XRT) \citep{kosugi} to study a number of bright points, using filter ratios to determine that their temperatures ranged between 1.1--3.4 MK. Another study by \cite{doschek} used EUV data from the \emph{Hinode} Extreme Ultra-violet Imaging Spectrometer, finding that the bright points reached maximum temperatures between 2--3 MK. \cite{alexander} studied the evolution of a single bright point, using both SXR and EUV data. This study found that throughout the 13 hours of observation, the bright point was almost isothermal, with an average temperature of 1.3 MK. The Hinode/XRT time profile for this bright point showed a steady increase in intensity for the first two hours of observation, followed by several spikes (which the authors speculated were likely due to heating or reconnection events) until it began to decay.
\par
Bright points can result from bipolar flux emergence. When this process occurs, new magnetic flux emerges to create an emerging flux region (EFR), which may subsequently evolve into a small-scale bright point, an example of which was investigated by \cite{kontogiannis}. This study tracked the evolution in Hinode/XRT of a bright point associated with an EFR, finding that the bright point exhibited a continuous increase in emission for $\sim$ 1.5 hours before it began to fade. However, rather than a bright point, an EFR may instead evolve into a large-scale active region if the emergence continues \citep{vandriel}.
\par
HXR observations of bright points have been made difficult by the faintness of this emission from these sources, and the lack of a solar dedicated instrument able to observe them individually. Previous HXR studies of the quiet Sun \citep{rhessi1,rhessi2} have been performed using data from the \emph{Reuven Ramaty High-Energy Solar Spectroscopic Imager} (RHESSI) \citep{lin}. However, as RHESSI was designed for the observation of bright sources, such as large flare events, only upper limits on the HXR emission from the whole solar disk were obtained. The authors noted that using a sensitive HXR focusing telescope would allow a more detailed study of the energy release mechanisms occurring in quiet Sun features. A more recent paper by \cite{buitrago} also constrains the quiet Sun HXR emission in the 5--10 keV range using data from the \emph{Focusing Optics X-ray Solar Imager} (FOXSI) sounding rocket. This study found similar upper limits to \cite{rhessi2}, though using only  minutes' worth of data, compared to the days that were required with RHESSI.
\par
The \emph{Nuclear Spectroscopic Telescope ARray} (NuSTAR) \citep{harrison} is a HXR focusing telescope which is capable of being pointed at the Sun to provide sensitive observations of faint solar sources \citep{bwgref}. Since 2014, there have been a number of NuSTAR solar observing campaigns\footnote{Summary of NuSTAR solar observations can be found at \url{https://ianan.github.io/nsigh_all/}}. Much of the work on the NuSTAR solar observations has focused on active region microflares \citep{wright,glesener1,hannah2019,cooper1,duncan,cooper2}, with \cite{glesener2} reporting for the first time on non-thermal emission detected in a microflare observed with NuSTAR. \cite{kuhar} presented work on quiet Sun flares observed by NuSTAR outside of an active region, finding that their temperatures ranged from 3.2--4.1 MK.
\par
NuSTAR's use of focusing optics means that it can directly image very faint HXR sources on the quiet Sun, and perform spectroscopy on regions of interest. During the recent solar minimum between cycles 24 and 25 (2018--2020), when the solar disk was free of active regions, NuSTAR was used to observe the Sun on a number of occasions, providing several bright points and other quiet Sun phenomena to study. The recent solar minimum combined with NuSTAR's sensitivity has provided a unique opportunity to study the HXR emission from these features and investigate their contribution to the heating of the solar atmosphere by searching for the presence of a high temperature ($>$5MK) or non-thermal component due to the presence of accelerated electrons.
\par
Here, we present the first survey of small features in the quiet Sun observed in HXRs with NuSTAR. We present analysis of several features from the 28 September 2018 full-disk mosaics, the first of the NuSTAR quiet Sun observations, including the first HXR imaging spectroscopy of such features. We include in our analysis EUV data from the \emph{Solar Dynamics Observatory's} Atmospheric Imaging Assembly (SDO/AIA) \citep{lemen} and SXR data from Hinode/XRT. An overview of this observation is presented in section \ref{overview}. The methods used to analyse the quiet Sun features are detailed in Section \ref{methods}. The detailed analysis of an EFR, X-ray bright points, a jet and bright limb source are presented in Sections \ref{efr}, \ref{bps}, \ref{jet} and \ref{limb}, respectively. A comparison of the thermal properties of these features are discussed in Section \ref{thermcomp}.

\section{Overview of Observation}
\label{overview}

\begin{figure}    
   \centerline{\includegraphics[width=1.0\textwidth,clip=]{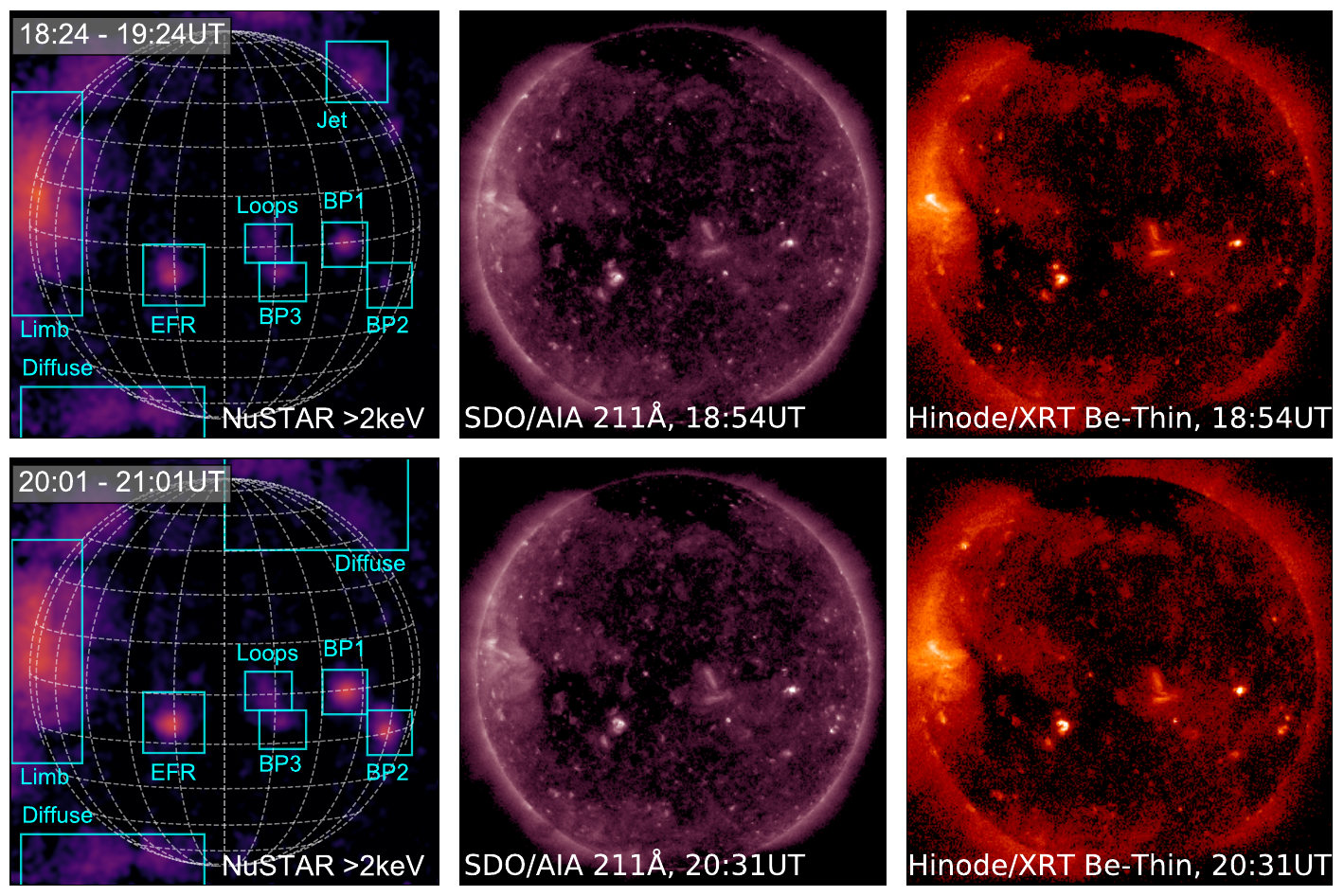}
   }
\caption{Full disk images for NuSTAR, SDO/AIA 211 Å and Hinode/XRT Be-thin for the two NuSTAR orbits of observation from 28 September 2018. The SDO/AIA and Hinode/XRT images are taken from the mid-times of each of the NuSTAR orbits.
        }
   \label{fig:fullmaps}
   \end{figure}

On 28 September 2018, NuSTAR observed the quiet Sun for two orbits (18:25 -- 19:25~UT and 20:01 -- 21:01~UT), producing a full-disk mosaic for each. As each of NuSTAR's two focal plane modules (FPMA and FPMB) cover approximately the same $12'$ $\times$ $12'$  FOV \citep{harrison}, multiple pointings are required to build up an image of the full-disk. These are formed over an orbit ($\sim$ an hour), comprised of 25 pointings each of duration $\sim$ 100 s, in a 5 $\times$ 5 grid pattern. The first pointing, P1, is in the top left corner, with the subsequent pointings shifting to the right until P5. P6 is shifted down from P5, with the following four pointings being increasingly shifted to the left. This pattern continues until P25, which is in the bottom right corner of the mosaic. The pointings overlap with each other such that a feature on the disk can be captured up to four times within a single orbit.
\par
For the times of these two orbits, there is also full-disk data available from SDO/AIA and Hinode/XRT (though there is a gap in the data from Hinode/XRT between 19:13--19:38 UT). The NuSTAR mosaics, and the SDO/AIA 211 Å and Hinode/XRT Be-thin full-disk images from the mid-times of both NuSTAR orbits, are shown in Figure \ref{fig:fullmaps}. NuSTAR is a photon-counting detector, and the data list the photons' properties, including time of detection, energy, and position on the detector. These NuSTAR full-disk maps are constructed by correcting each pointing for livetime individually and aligning the NuSTAR image with SDO/AIA, and then summing the corrected counts. 
\par
Because of the way in which the mosaic tiles overlap, the disk is sampled more times than the limb (which is captured in only one or two pointings, compared to four pointings on the disk). An additional correction has been applied per detector quadrant to these mosaics to account for this by normalising by the number of times a given region has been sampled over the whole mosaic. Note that these NuSTAR mosaics, as well as all of the NuSTAR images shown here, have been smoothed with a Gaussian filter.

\par
There are several features present in the NuSTAR images shown in Figure \ref{fig:fullmaps}, with corresponding sources appearing also in SDO/AIA and Hinode/XRT. 

\begin{itemize}
	\item One feature bright in both orbits is an EFR (just left of disk centre in Figure \ref{fig:fullmaps}) which went on to become the NOAA active region 12723$\beta$ a few days after this observation. Its properties and evolution are detailed in Section \ref{efr}.
	\item A number of X-ray bright points are identified, two appearing brightly in the second NuSTAR orbit (first panel of the bottom row in Figure \ref{fig:fullmaps} - which we label BP1 and BP2), and well as a fainter one (we label BP3) which is nearby some quiet Sun loops. Analysis of these features is detailed in Section \ref{bps}.
	\item In just the first NuSTAR orbit (first panel of the top row in Figure \ref{fig:fullmaps}), there is a faint source that appears in the top right corner of the mosaic, which can be seen to be a jet in SDO/AIA (see Section \ref{jet}).
	\item The brightest source in the NuSTAR images is the large region at the east limb, seen both NuSTAR orbits. Between the NuSTAR detectors there are chip gaps, and any photons that land on these gaps are not detected. This source is bright and extended enough to cross over the chip gaps between NuSTAR's detector quadrants, with these detector gaps are visible in Figure \ref{fig:fullmaps}. The bright loops in this region, clearer in the SDO/AIA and XRT images, are the remains of a decayed active region (DAR), details and analysis given in Section \ref{limb}.
\end{itemize}

\section{Analysis Methods}
\label{methods}
\subsection{Fitting the NuSTAR HXR Spectra}
\label{spectralfitting}

NuSTAR is an imaging spectrometer, allowing the X-ray spectra of these quiet Sun features to be fit to investigate their properties. The spectral fitting of the features presented here was done using XSPEC \citep{arnaud}. In order to preform spectral fitting for a given source, the spectrum and Spectral Response Matrix (SRM) - via the the Response Matrix and Ancillary Response Files (RMF and ARF) - were obtained for a circular region enclosing the source (in this study, all with radii $> 40''$) using the NuSTAR Data Analysis Software. As the features investigated here produce a low number of counts, Cash statistics \citep{cash}, the maximum likelihood-based statistic for Poisson data, were used for the fitting. For the APEC thermal model used in XSPEC, coronal abundances were assumed. 

For all of the NuSTAR quiet Sun features investigated here, the HXR emission is faint (resulting in high livetimes between 67--92 \%). Although NuSTAR has a high sensitivity its detector throughput is limited to 400 counts/s/FPM. This limited throughput combined with HXR spectra that are sharply falling off with increasing photon energy (whether due to thermal or non-thermal continuum sources) results in the low energy counts dominating. This observed over the short duration of the mosaic pointing times ($\sim$ 100s), means only noisy spectra are observed with few, or no counts, above a few keV. These noisy spectra over a limited energy range are tricky to fit. Our study is helped by the recent update to NuSTAR's calibration which makes it possible to fit down to 2.2 keV \citep{madsen}, whereas previously only down to 2.5 keV was recommended for solar observations \citep{bwgref}.

Our noisy spectra can be improved by simultaneously fitting the FPMA and FPMB spectra, introducing a multiplicative constant to the fits to account for any systematic difference between the responses of the two telescopes. This constant is a fit parameter for FPMB (so relative to FPMA value) which varies depending on this systematic uncertainty and where the source lies on the detector. As most of the features presented in this paper are very faint, using the simultaneously fitted two FPMs' spectra is not enough. Fortunately, all of the features were captured in more than one pointing due to the overlapping mosaic tiles, so the spectra from multiple FPMs and pointings can be simultaneously fitted to improve the signal-to-noise ratio. However, it is important to find a balance between obtaining a good fit, and being able to investigate how a source evolves over time.

\subsection{Reconstructing Differential Emission Measures}
\label{dems}

To investigate the multi-thermal nature of a given feature, a differential emission measure can be recovered by combining data from NuSTAR, Hinode/XRT, and the six SDO/AIA optically thin coronal-temperature channels. As with the spectral fitting, we can consider multiple pointings jointly by averaging both the NuSTAR data values and responses over the pointings. The Python version of the regularized inversion approach of \cite{hannahdem} was used to reconstruct the DEMs for the features, weighted using the minimum of the EM loci curves (the data divided by the corresponding temperature response function, giving the maximum possible emission at each temperature). 
\par
The fluxes for SDO/AIA and Hinode/XRT are obtained from an image from each channel averaged over the relevant NuSTAR pointing time. A systematic error of 20$\%$ is assigned to the SDO/AIA, Hinode/XRT, and NuSTAR fluxes in order to account for uncertainties in their temperature responses. When calculated, the photon shot noise for SDO/AIA and Hinode/XRT was found to be negligible for these features as the data values were calculated over relatively large regions. However, in the case of NuSTAR, the calculated shot noise was not negligible, and therefore is added in quadrature with the 20$\%$ systematic error for all DEMs shown here. For the DEM calculation, the NuSTAR data is split into two energy bands: 2.2--2.6 keV and 2.6--3.6 keV. The SDO/AIA and Hinode/XRT temperature responses were calculated using the standard Solarsoft routines from the instrument teams: \texttt{aia\_get\_response.pro} for SDO/AIA and \texttt{make\_xrt\_temp\_resp.pro} for Hinode/XRT. The NuSTAR response was calculated in Python\footnote{\url{https://github.com/ianan/nustar_sac/blob/master/python/ns_tresp.py}} using the spectral responses used for the spectral fitting.
\par
DEM analysis combining data from NuSTAR, SDO/AIA and Hinode/XRT has been done before by \cite{wright}. In this study, it was found that multiplying the Hinode/XRT responses by a factor of 2 before calculating the DEMs produced a solution with smaller residuals. This choice was made following the suggestion of previous authors who also found discrepancies when using Hinode/XRT data \citep{schmelz,testa,cheung}. Introducing this factor was similarly found to improve the DEM results here, and so was used for all of the DEMs presented in this paper.

\subsection{NuSTAR Non-Thermal Upper Limits}
\label{upperlimits}

The NuSTAR spectra of these quiet Sun features are well-fitted by an isothermal model, but as they are noisy or have no counts at higher energies (possibly due to NuSTAR's limited detector throughput) a weak non-thermal component could be present but undetected. We can determine an upper limit on the non-thermal emission that could be present and consistent with a null detection, following the approach of \cite{wright}. This is done by adding a non-thermal component to the thermal model obtained from the NuSTAR spectral fitting. This thick target non-thermal model depends on three parameters: the power-law index, $\delta$, the low-energy cutoff, $E_{c}$, and the total electron flux, $N_{N}$. For a chosen $\delta$, $E_{c}$ and $N_{N}$ value, the resulting non-thermal model and the fitted thermal model are folded through the NuSTAR response, producing model count spectra. From these, synthetic spectra are generated through a Monte Carlo process, randomly sampling the model count spectra for the total number of counts (calculated using the livetime and duration of the observation). For a range of different $\delta$ and $E_{c}$ combinations, $N_{N}$ is reduced until these synthetic spectra  lie within each others' Poisson errors between 2 and 4 keV, and there are $<$ 4 counts above 4 keV -- consistent with a null detection to 2$\sigma$ \citep{gehrels}. Because the spectra considered here are noisy, this test can be repeated multiple times (1000 times in the cases discussed later), to obtain more accurate results. As well as testing models with different $\delta$ values, the case of a mono-energetic beam of electrons, with an energy of $E_{c}$, can also be tested.  These simulations were done in Python using the thermal and thick-target models from the new solar X-ray fitting package\footnote{\url{https://github.com/sunpy/sunxspex/blob/master/sunxspex/sunxspex_fitting/photon_models_for_fitting.py}}.

The upper limit on $N_{N}$ can then be used to determine an upper limit on the power in the non-thermal distribution, via:
\begin{equation}
P(>E_{c}) = 1.6 \times 10^{-9}\frac{\delta-1}{\delta-2} N_{N}E_{c}  \quad     [\mbox{erg s}^{-1}]
\label{eq:power}
\end{equation}

This can then be multiplied by the duration of the observation and compared to the thermal energy, calculated as:

\begin{equation}
E_{th}=3k_{B}T\sqrt{EM\mbox{ }V} \quad [\mbox{erg}]
\label{eq:energy}
\end{equation}
where $k_{B}$ is the Boltzmann constant, V is the volume of the emitting plasma (found from the SDO/AIA or Hinode/XRT image, using source area $A^{3/2}$), and T and EM are the temperature and emission measure of the plasma (from the NuSTAR spectral fit), respectively \citep{hannah}. Note that this equation does not take into account a loop filling factor, which means that this is an upper limit on the thermal energy. This calculated thermal energy then provides a heating requirement for the source, which if less than the upper-limits of the non-thermal power, could be the produced by the accelerated electrons, like in larger flares.

\section{Emerging Flux Region} 
      \label{efr} 

The EFR, which later went on to become an active region, was observed by NuSTAR in both orbits. For this EFR, flux first begins to emerge just after 00:00 UT on 28 September. There is some cancellation between opposite polarities, and the positive and negative polarities then spread apart. This is the time during which this feature is observed by NuSTAR. The next day, beginning at $\sim$ 14:00 UT on 29 September, there is more intense flux emergence in this region, producing an active region.
\par

\begin{figure}    
   \centerline{\includegraphics[width=1.0\textwidth,clip=]{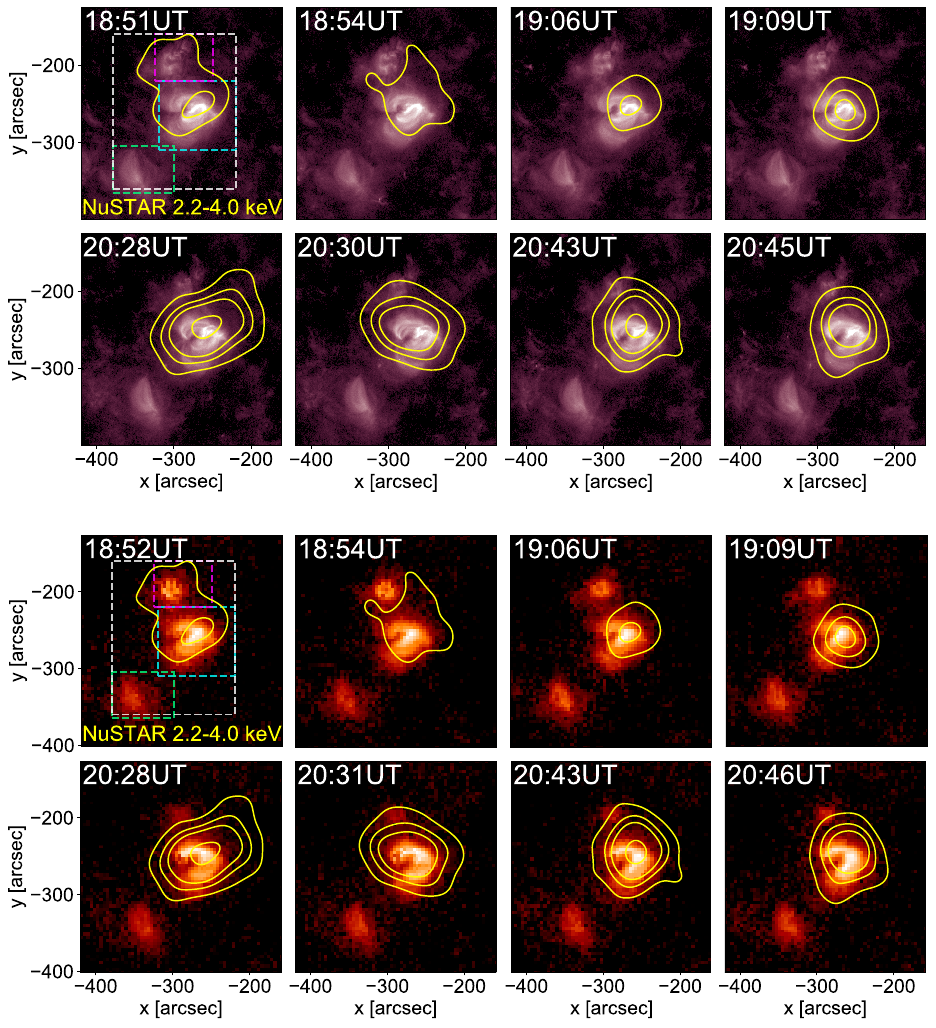}
              }
              \caption{SDO/AIA 211Å (top two rows) Hinode/XRT Be-thin (bottom two rows) images of the EFR from the mid-times of the eight NuSTAR pointings that capture the feature. Yellow contours represent aligned NuSTAR FPMA + FPMB 2.2--4.0 keV, with the contours plotted at the same levels in all panels (5, 10, 15, 30 $\times$ $10^{-4}$ counts s$^{-1}$). For each instrument, the top row shows orbit 1 and the bottom shows orbit 2, with P12, 13, 18, 19 arranged from left to right. The lightcurves in Figure \ref{fig:efrlightcurves} were calculated for the dashed boxes.}
   \label{fig:efrcontours}
   \end{figure}

In each NuSTAR orbit, the EFR was captured in four pointings, split into two pairs of consecutive pointings: 12 and 13, and 18 and 19. Capturing this feature four times in each orbit, for a total number of eight pointings over the whole observation, gives the opportunity to study its temporal evolution despite the short duration of the mosaic pointings. 
\par
Hinode/XRT Be-thin and SDO/AIA 211 Å images of the EFR are shown in Figure \ref{fig:efrcontours}. These images are from the mid-times of all of the mosaic pointings where NuSTAR captured the EFR. The NuSTAR 2.2--4.0 keV contours, aligned with SDO/AIA, are also plotted in this figure to show the HXR evolution of the feature. These contours indicate that the feature is generally brighter in the second orbit than the first. In the Hinode/XRT and SDO/AIA images, the EFR appears to be comprised of three separate regions. It is clear that the majority of the NuSTAR emission originates from the bright central region ($-250'', -250''$), particularly in the second orbit when the feature has brightened. However, there does appear to be a contribution from the upper region ($-300'', -200''$) to the NuSTAR emission in P12 of the first orbit, as shown in the top left panel for each instrument of Figure \ref{fig:efrcontours}. The lower region ($-350'', -350''$) does not appear to contribute significantly to the NuSTAR emission. This is expected in P12 and 13 of each orbit, as this region would be outside NuSTAR's FOV. In P18 and 19, this feature would lie close to a detector gap, which could explain the lack of NuSTAR emission from this region.
\par

\begin{figure}    
   \centerline{\includegraphics[width=1.0\textwidth,clip=]{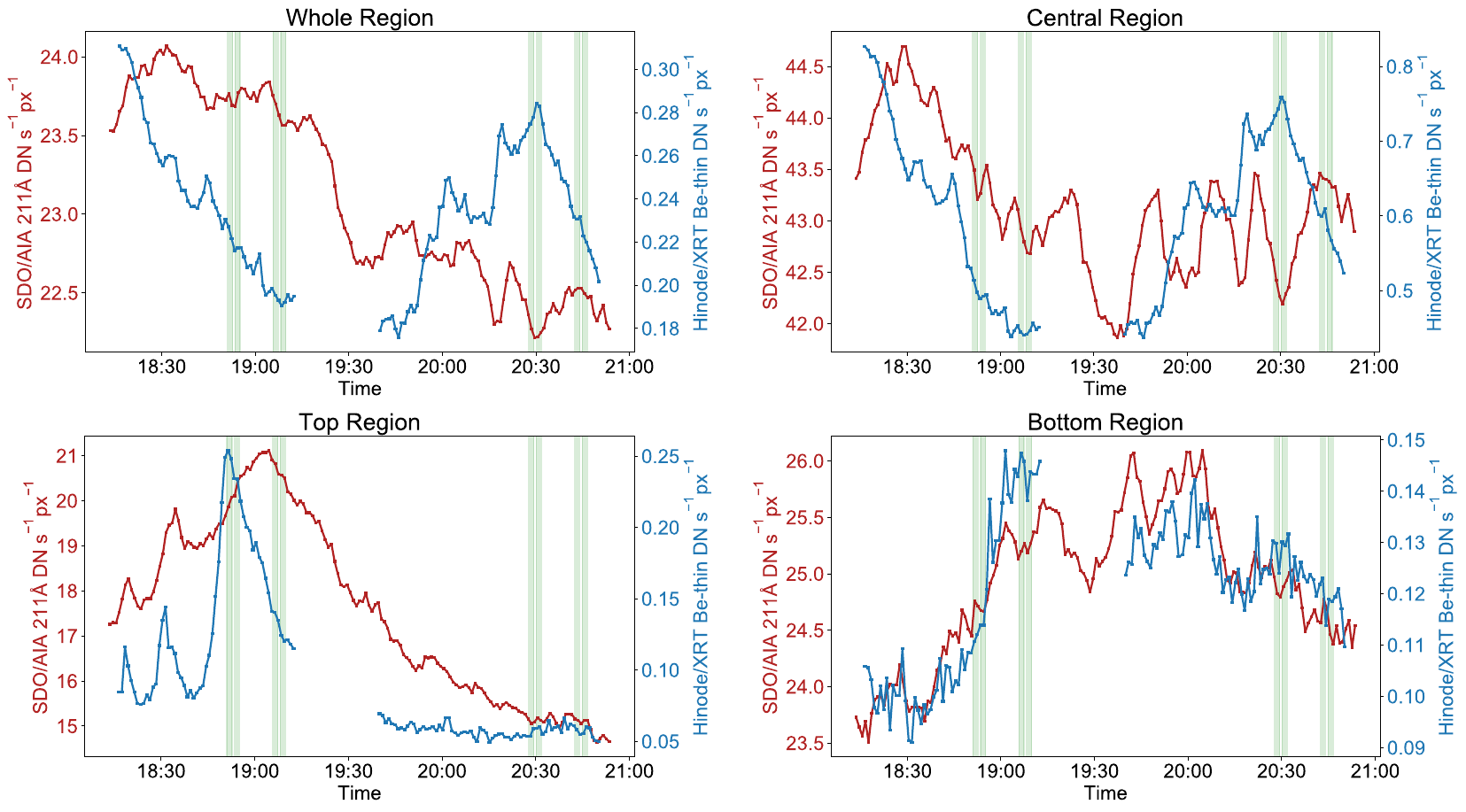}
              }           
\caption{Hinode/XRT Be-thin (blue) and SDO/AIA 211 Å (red) lightcurves for the whole EFR, as well as the three separate regions within it. The green shaded areas indicate the times of the eight NuSTAR mosaic pointings that captured the EFR.
        }
   \label{fig:efrlightcurves}
   \end{figure}

\par
The Hinode/XRT Be-thin and SDO/AIA 211Å lightcurves for the EFR are plotted in Figure \ref{fig:efrlightcurves}, for each of the three regions separately (the boxes shown in the top left panel in Figure \ref{fig:efrcontours}) and combined. These lightcurves confirm that the central region is the brightest of the three in both SDO/AIA and Hinode/XRT, and the greatest contributor to the emission from the EFR as a whole, as shown in the top row of Figure \ref{fig:efrlightcurves}. In the first orbit, both lightcurves for this region decrease between the times of the NuSTAR pointings (18:53 UT and 19:07 UT, shown by the green shaded regions in Figure \ref{fig:efrlightcurves}). Interestingly, between the two NuSTAR pointing times in the second orbit (20:29 UT and 20:44 UT) the Hinode/XRT lightcurve decreases where the SDO/AIA 211 Å lightcurve increases. However, the change in the SDO/AIA 211 Å lightcurve is relatively much smaller than the change in Hinode/XRT.
\par
The Hinode/XRT lightcurve for the upper region shows a sharp peak in brightness during P12 of the first NuSTAR orbit, before falling off. This is consistent with the NuSTAR contours in the top left panel for each instrument in Figure \ref{fig:efrcontours}, where the upper region is significantly contributing to the NuSTAR emission in P12 of orbit 1. The peak in the Hinode/XRT lightcurve followed later by the peak in the SDO/AIA 211 Å lightcurve for this region suggests a heating of material followed by cooling. 
\par
The lower region is outside NuSTAR's FOV in P12 and 13, and would be positioned near or on a detector gap in P18 and 19. However, the lightcurves confirm that this feature is relatively very faint in Hinode/XRT in comparison to the central region, and therefore would be unlikely to contribute significantly to the NuSTAR emission from the EFR.

\subsection{NuSTAR Spectral Analysis}\label{nsspefr}
  
\begin{figure}    
\centering 
\includegraphics[width=1.0\textwidth,clip=]{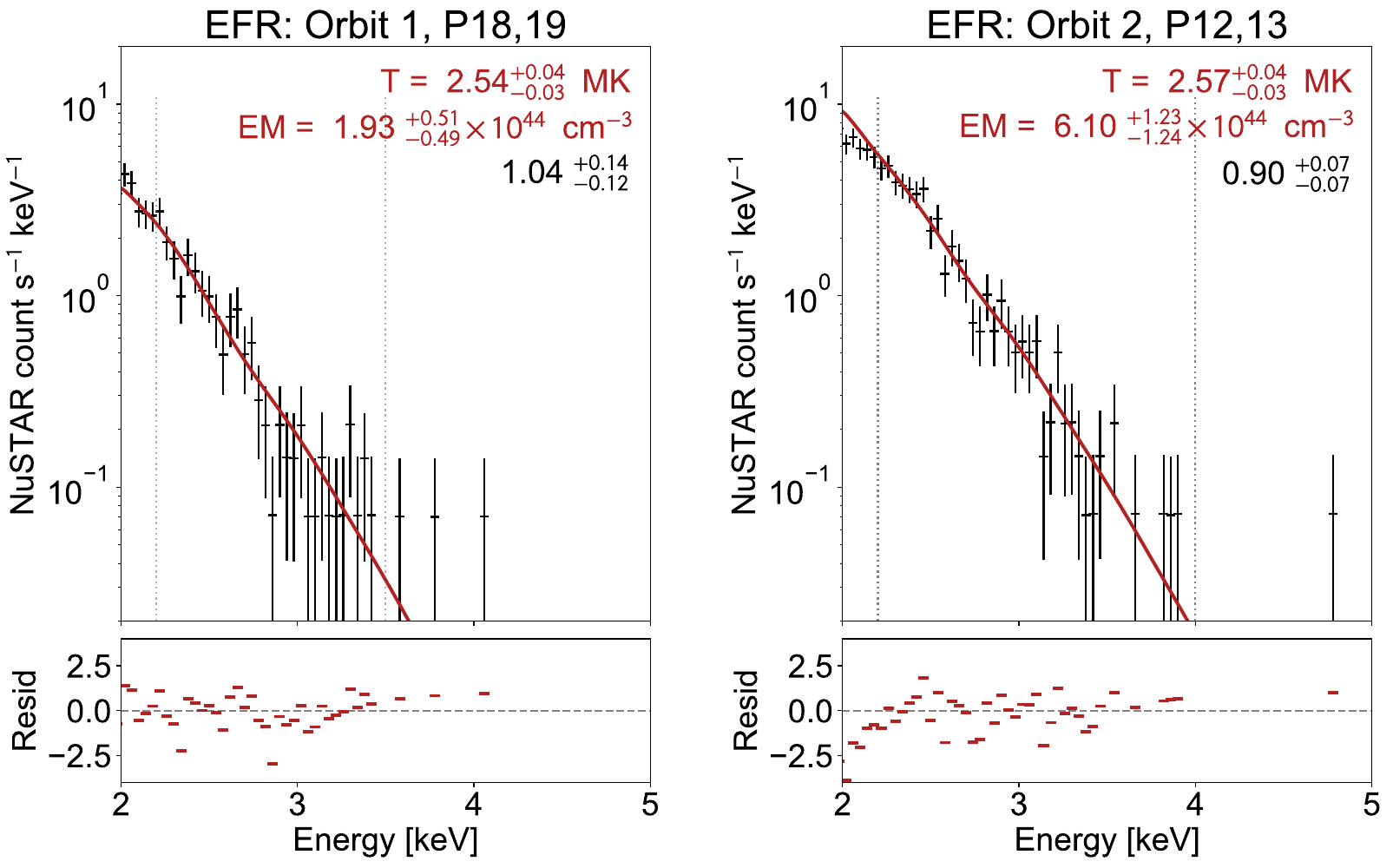}
\caption{NuSTAR fitted spectra for the EFR, (left) orbit 1 P18, P19 and (right) orbit 2 P12, 13. The red line and numbers indicate the fitted thermal model, dotted lines the fitting range, and the black number is the multiplicative constant to account for systematic differences between FPMA and FPMB.}
   \label{fig:efrspec}
   \end{figure}

\par
Using the approach detailed in Section \ref{spectralfitting}, we fit an isothermal model  to the NuSTAR spectra of the EFR. Multiple pointings were simultaneously fit over to reduce the noise, but it was sufficient just to use consecutive pointings (P12,13 and P18,19) so that we could still investigate the time evolution. The spectra for orbit 1 P18,19 and orbit 2 P12,13 are shown in Figure \ref{fig:efrspec}, and a summary of the fitting results for the EFR (in comparison with all other features investigated) is given in Table \ref{specresults}. The spectral fits from all four times give a reasonably constant temperature of $\sim$ 2.5 MK, and emission measures ranging between 1.9 and 6.1 $\times$ $10^{44}$ cm$^{-3}$. The fits do suggest a slight increase in temperature of the EFR over the two orbits, from 2.54 MK to 2.63 MK. However, taking into account the uncertainties on these temperatures, this increase is not statistically significant. The emission measure from the spectral fits decreases from 3.42 to 1.93 $\times$ $10^{44}$ cm$^{-3}$ between P12,13 and P18,19 in the first orbit. It then increases up to 6.10 $\times$ $10^{44}$ cm$^{-3}$ in P12,13 in the second orbit, before falling to 2.40 $\times$ $10^{44}$ cm$^{-3}$ for P18,19. 
\par
This matches the behaviour in the Hinode/XRT lightcurve for the central region in Figure \ref{fig:efrlightcurves}. In both orbits, the NuSTAR fit temperature remains approximately constant while the emission measure decreases. This is in agreement with the decreases in the Hinode/XRT lightcurve between the NuSTAR pointing times in both orbits, and the highest NuSTAR emission measure corresponds to the highest peak in Hinode/XRT at 20:30 UT. This similar behaviour is expected as Hinode/XRT and NuSTAR should be observing emission at approximately the same temperatures.

\subsection{Differential Emission Measures}
\label{efrdems}

In the EFR lightcurves in Figure \ref{fig:efrlightcurves}, at around 20:30 UT (NuSTAR orbit 2, P12 and 13), there is a peak in the Hinode/XRT Be-thin lightcurve that coincides with a minimum in the SDO/AIA 211 Å lightcurve. Later, at around 20:45 UT (NuSTAR orbit 2, P18 and 19) there is increased SDO/AIA 211Å emission but decreased Hinode/XRT Be-thin emission. As Hinode/XRT is sensitive to higher temperature emission than SDO/AIA 211 Å, this suggests that there is more higher temperature emission present at 20:30 UT than at 20:45 UT. In order to confirm this we perform differential emission measure (DEM) analysis using the method outlined in Section \ref{dems}.
\par

\begin{figure}    
\centering
   \includegraphics[width=0.98\textwidth,clip=]{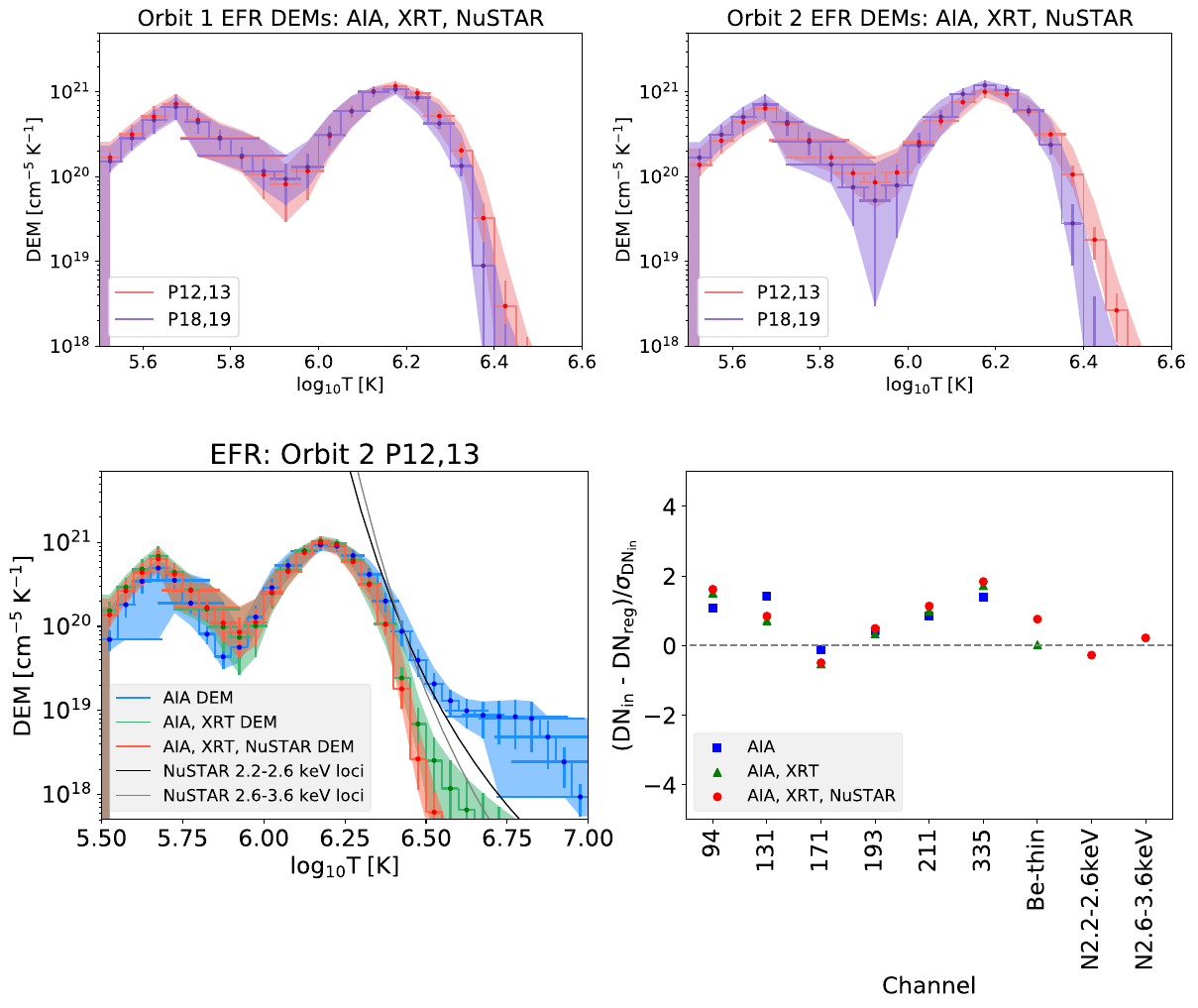}                            
\caption{(Top row) EFR DEMs for NuSTAR orbit 1 (left) and orbit 2 (right), combining P12 and 13 (peach) and P18 and 19 (purple), the shaded area representing the uncertainty. These DEMs are reconstructed using data from NuSTAR, Hinode/XRT and SDO/AIA. (Bottom row) For the EFR during orbit 2 P12 and 13, the DEM has been reconstructed using SDO/AIA only (blue), SDO/AIA and Hinode/XRT (green) and NuSTAR, Hinode/XRT and SDO/AIA (red), with the NuSTAR EM loci curves for comparison. The corresponding normalised residuals in data space are shown in the bottom right panel.}
   \label{fig:efrdems}
   \end{figure}

A comparison between the DEMs calculated for P12 and 13, and P18 and 19 is shown for each orbit in the top row of Figure \ref{fig:efrdems}. In orbit 1, between these two times, the emission in both the Hinode/XRT Be-thin and SDO/AIA 211 Å lightcurves in Figure \ref{fig:efrlightcurves} decreases, though the change in brightness is smaller than the 20$\%$ systematic error used in the DEM calculation in both cases. This is reflected in the two DEMs for these times, as the DEM for P12 and 13 is higher than the one for P18 and 19 for $\log_{10} T\gtrsim$ 6.2, though they are not significantly different when taking into account the error bars. 
\par
In the case of orbit 2, the Hinode/XRT lightcurve shows a peak at around the time of NuSTAR P12 and 13, before decreasing for NuSTAR P18 and 19. Between these two times, the decrease in Hinode/XRT emission and corresponding increase in SDO/AIA 211 Å emission suggests that there is hotter material present at the earlier time. The change in Hinode/XRT is higher than the 20$\%$ systematic errors, whereas the change in 211 Å is again very small. From the calculated DEMs, there is more emission above $\log_{10} T\gtrsim$ 6.3 at the earlier time, in agreement with the fall-off in the Hinode/XRT lightcurve. The DEM for P18 and 19 is higher than that for P12 and 13 for 6.1 $\lesssim \log_{10} T\lesssim 6.3$, though there is no difference between the DEMs at these temperatures outwith the error regions. Material at these temperatures could be responsible for the peak in the SDO/AIA 211Å at the later time.
\par
 Adding X-ray data to a DEM calculation using SDO/AIA is important to constrain  the higher temperature emission, as shown in the bottom row of Figure \ref{fig:efrdems}. In this figure, the intersection of the NuSTAR EM loci curves, obtained by dividing the data by the NuSTAR temperature response in each energy bin, is consistent with the T and EM values from the NuSTAR spectral fit, as expected. All three DEMs are similar for $\log_{10} T\lesssim 6.4$ but demonstrate that the addition of Hinode/XRT helps to constrain the DEM at temperatures higher than this, and adding NuSTAR strengthens this constraint. From including the X-ray data rather than relying on SDO/AIA alone, it is clear that there is virtually no emission above 4 MK here. Previous studies looking at non-flaring active regions using different DEM approaches with SDO/AIA data also produced erroneous higher temperature DEM components, removed when HXR data was included \citep{schmelz2,reale}. A previous study of a quiet Sun EFR by \cite{kontogiannis} also included DEM analysis, using data from Hinode/EIS, finding a similar magnitude peak in the DEM at $\log_{10} T\sim$ 6.1 to those shown in Figure \ref{fig:efrdems}. As \cite{kontogiannis} only used Hinode/EIS, their DEMs were not well constrained at higher temperatures. 
\par

\subsection{NuSTAR Non-thermal Upper Limits}
Using the approach detailed in  Section \ref{upperlimits}, the upper limits on any non-thermal emission present in the EFR were calculated and compared to the required heating power obtained from the thermal energy of the plasma. This calculation was done for the peak time in the Hinode/XRT lightcurve, corresponding to NuSTAR orbit 2 P12 and 13. The central region of the EFR is $\sim$ $35''$ square  (see Figure \ref{fig:efrcontours}) and, using $V\sim A^{3/2}$, has a volume of 1.65 $\times$ $10^{28}$ cm$^{3}$. Using equation \ref{eq:energy} and the NuSTAR spectral fit values (see Figure \ref{fig:efrspec}), we find a thermal energy of 3.37 $\times$ $10^{27}$ erg, and hence, by dividing by the NuSTAR observation time (246~s),
a heating power of 1.37 $\times$ $10^{25}$ erg s$^{-1}$.
\par
 We find that all of the upper limits on the non-thermal power were smaller than the heating requirement. The area used here may have been an over-estimate, and making this  smaller would lower the heating requirement, but only by a small factor -- this would still be an order of magnitude larger than the upper limits on the non-thermal power. Therefore, it can be concluded that, in the case of the EFR, if any non-thermal component is present, it is not responsible for the observed heating.

\section{Bright Points}
\label{bps}

\begin{figure}    
   \centering
   \includegraphics[width=1.0\textwidth,clip=]{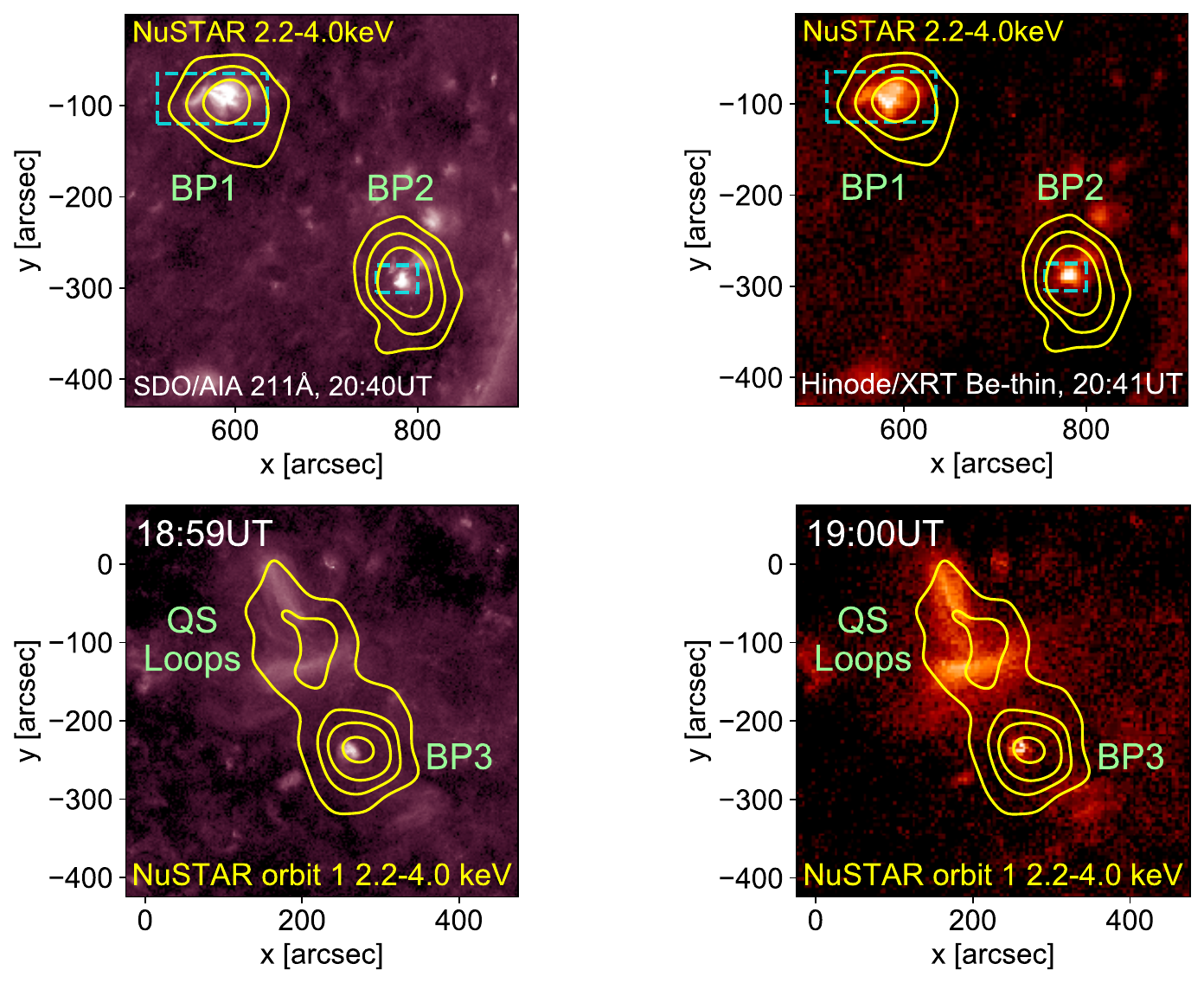}  
\caption{(Top row) SDO/AIA 211 Å (left) and Hinode/XRT Be-thin (right) images of the two bright points from 20:41 UT, coinciding with NuSTAR orbit 2 P17. Yellow contours represent NuSTAR 2.2--4.0 keV (contour levels are 5, 10, 20 $\times$ $10^{-4}$ counts s$^{-1}$), aligned with SDO/AIA. The dashed boxes indicate regions used for obtaining the lightcurves in Figure \ref{fig:bplightcurves}. (Bottom row) SDO/AIA 211 Å (left) and Hinode/XRT Be-thin (right) images of the QS loops and bright point (BP3) near disk centre from orbit 1. The aligned 2.2--4.0 keV NuSTAR contours are summed over P13, 14, 17, and 18 of orbit 1, and are plotted at  4, 7, 10, 15 $\times$ $10^{-4}$ counts s$^{-1}$.} 
   \label{fig:bpcontours}
   \end{figure}

Three bright points were identified in the NuSTAR observations and confirmed with Hinode/XRT and SDO/AIA, as was shown in Figure \ref{fig:fullmaps}. Zoomed-in maps of these bright points, with images in Hinode/XRT and SDO/AIA 211 Å and over-plotted NuSTAR contours, are shown in Figure \ref{fig:bpcontours}. The top row shows bright points labelled BP1 and BP2 from NuSTAR orbit 2. In both 211 Å and Hinode/XRT, BP2 is a more compact feature than BP1, but it is brighter in NuSTAR at this time. BP1 is observed with NuSTAR in P14, 15, 16, and 17 of both orbits. BP2 is close to BP1, and lies in a region that is captured by NuSTAR also in P14, 15, 16, and 17. However, this feature is extremely faint in the first orbit, making it unusable for spectroscopy. Though it is present in all pointings in orbit 2, BP2 is located over the edge of the detector in P14 and 15, and is therefore only well observed in P16 and 17.
In the bottom row of Figure \ref{fig:bpcontours}, a fainter bright point, labelled BP3, is shown near to some larger loops in the quiet Sun (labelled QS loops). Each of these features were captured in P13, 14, 17, and 18 in both of the NuSTAR orbits.

\begin{figure}    
\centerline{\includegraphics[width=1.0\textwidth,clip=]{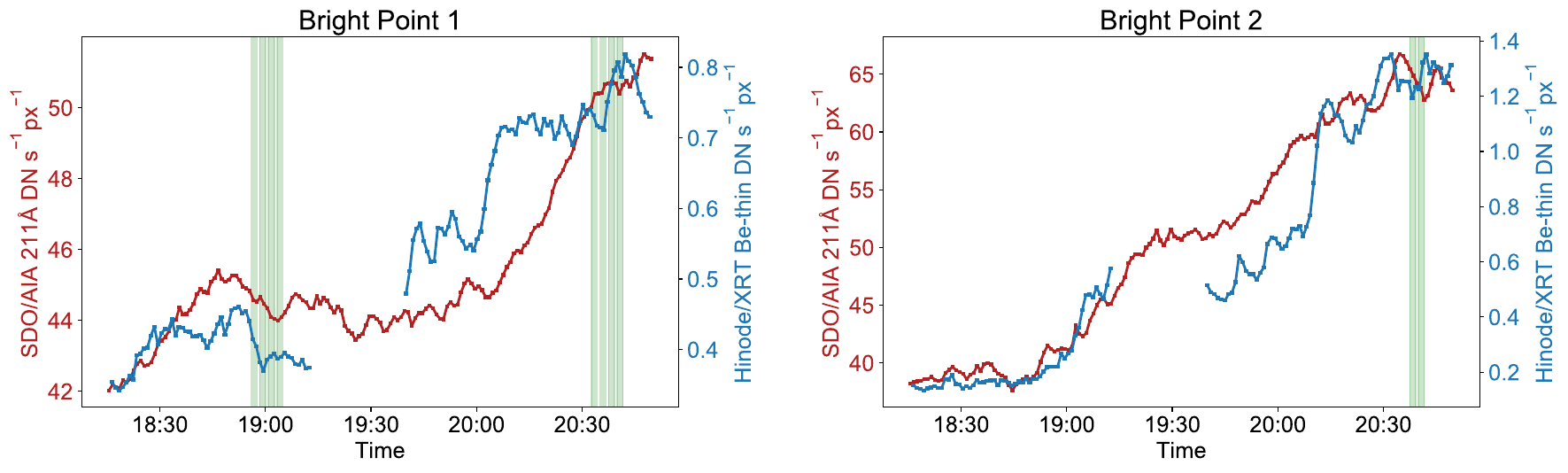}
              }
\caption{SDO/AIA 211 Å (red) and Hinode/XRT (blue) Be-thin lightcurves for BP1 (left) and BP2 (right). The green shaded areas indicate the times of the NuSTAR pointings suitable for spectroscopy.
        }
   \label{fig:bplightcurves}
   \end{figure}

The SDO/AIA 211Å and Hinode/XRT Be-thin lightcurves for the BP1 and BP2 are shown in Figure \ref{fig:bplightcurves}, calculated for the boxes shown in Figure \ref{fig:bpcontours}. In the case of BP1, the lightcurves indicate an increase in brightness in both channels, peaking just before the NuSTAR pointings in the first orbit, and then they continue to increase in brightness until the pointings in the second orbit. The two lightcurves for BP2 also show increasing brightness throughout both orbits of NuSTAR observation. This behaviour explains why BP2 is not observed by NuSTAR in the first orbit; it is not yet producing sufficiently bright emission to be detected by NuSTAR. SDO/AIA 211Å and Hinode/XRT lightcurves for the QS loops and for BP3 (not shown here) show only small changes in brightness for these features between the two orbits.

\subsection{NuSTAR Spectral Analysis}

 \begin{figure}    
    \centering
    \includegraphics[width=1.0\textwidth,clip=]{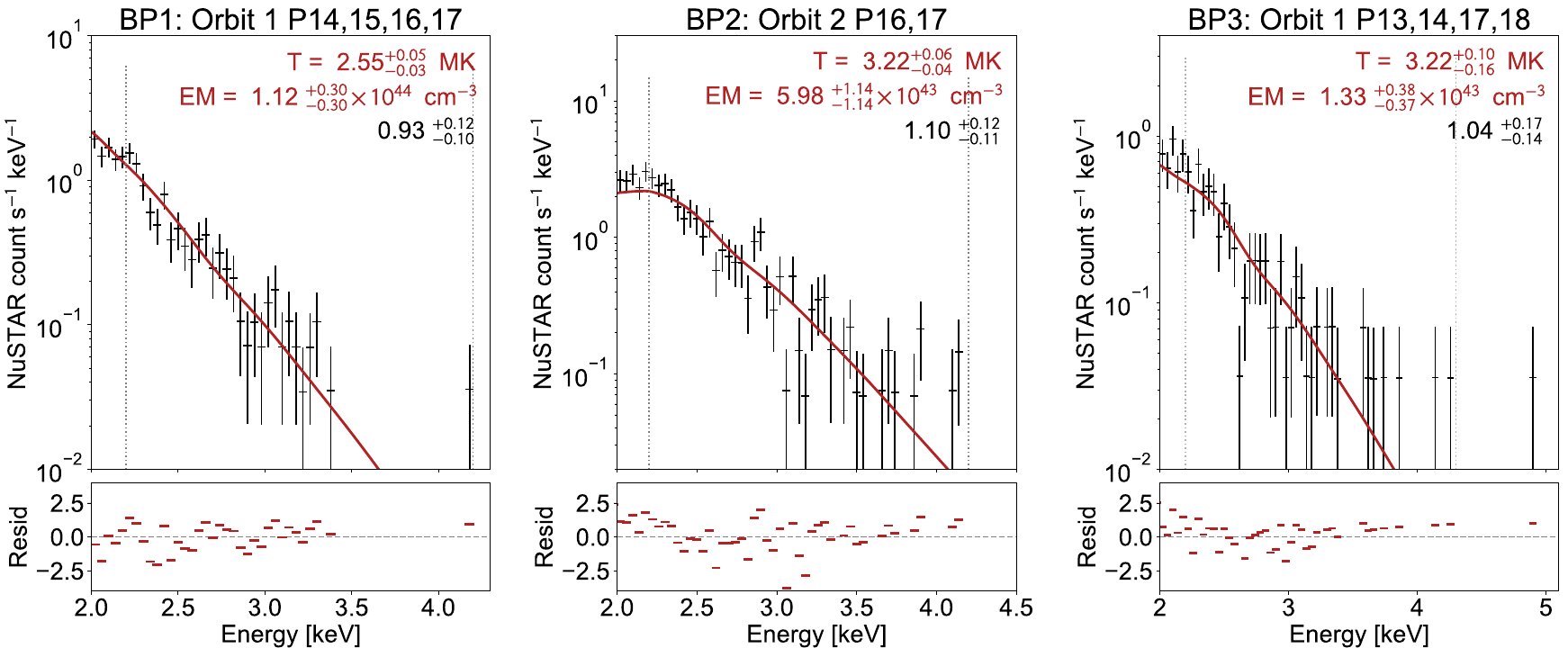}
    
 \caption{NuSTAR spectral fits for BP1 (left, during orbit 1 combining P14, 15, 16, 17), BP2 (middle, orbit 2 combining P16, 17) and BP3 (right, orbit 1 combining P13, 14, 17, 18) the red line and text indicating the thermal model and parameters found. All used FPMA and FPMB spectra, with the multiplicative constant introduced to account for systematic differences between FPMA and FPMB given by the black text. The energy range fitted is shown by the vertical dotted lines.}
    \label{fig:bpspec}
    \end{figure}

Figure \ref{fig:bpspec} shows example NuSTAR spectra for BP1, BP2 and BP3 fitted with a thermal model. For each bright point (except BP2), and the QS loops, spectra were obtained and fitted for both orbits, and all the fitted thermal parameters are given in Table \ref{specresults}.

For BP1, the spectral fits give a temperature of $\sim$ 2.5 MK in both orbits but with emission measure increasing from 1.12 $\times$ $10^{44}$ cm$^{-3}$ in orbit 1 (as shown in left panel of Figure \ref{fig:bpspec})  to 4.07 $\times$ $10^{44}$ cm$^{-3}$ in orbit 2, matching the behaviour seen in the Hinode/XRT and SDO/AIA lightcurves (Figure \ref{fig:bplightcurves}). Although, NuSTAR gives an EM increase by a factor of $\sim$ 4, while the feature brightens by a factor of $\sim$ 1.2 in SDO/AIA 211 Å and $\sim$ 2 in Hinode/XRT Be-thin. This could be due to both of these channels, 211Å in particular, being sensitive to cooler material than NuSTAR, meaning that this brightening might be occurring slightly higher temperatures, and is less significant to the overall emission.

As BP2 was only well observed during NuSTAR orbit 2 P16 and 17, only this spectrum could be fitted, finding 3.22 MK and 5.98 $\times$ $10^{43}$ cm$^{-3}$,  respectively (shown middle panel of Figure \ref{fig:bpspec}). BP3 is fainter but was well observed over both NuSTAR orbits, the fitted spectrum in orbit 1 giving 3.22 MK and 1.33 $\times$ $10^{43}$ cm$^{-3}$ (shown right panel of Figure \ref{fig:bpspec}), and 2.56 MK and 5.10 $\times$ $10^{43}$ cm$^{-3}$ during orbit 2. The QS loops were also well observed over both NuSTAR orbits, the fitted spectra giving cooler similar or slightly cooler temperatures (2.07--2.51 MK) than the bright points (see Table \ref{specresults}). All the BP (and QS loop) spectra are well fitted with the isothermal model and, similarly to the previous examples, do not show any evidence of either a higher temperature or non-thermal component.

\subsection{Differential Emission Measures}

\begin{figure}    
\centering
\includegraphics[width=1.0\textwidth,clip=]{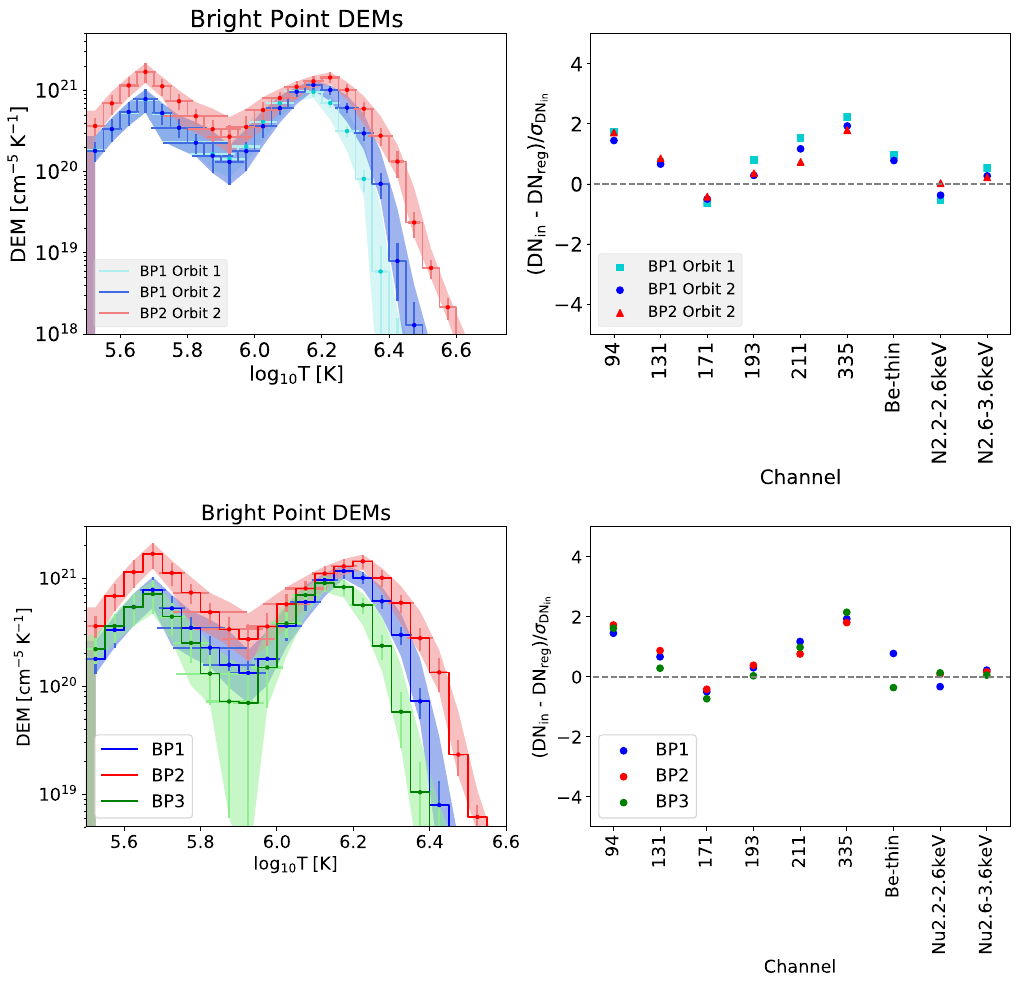}
\caption{(Top row) DEMs for for BP1 in orbits 1 (light blue) and 2 (dark blue) were reconstructed using data from SDO/AIA, Hinode/XRT, and NuSTAR, and for BP2 in orbit 2 (red) was reconstructed using data from only SDO/AIA and NuSTAR (due to saturation in Hinode/XRT Be-thin). (Bottom row) Comparison of the DEMs for BP1 and BP2 from orbit 2, and BP3 from orbit 1 when this feature is hottest.}
\label{fig:bpdems}
\end{figure}

Following the approach of section \ref{dems}, DEMs were reconstructed for the bright points and are shown in Figure \ref{fig:bpdems}. DEMs are shown for both orbits for BP1, but only in orbit 2 for BP2 and several of the Hinode/XRT Be-thin pixels which covered BP2 were saturated at the times of these pointings, so only SDO/AIA and NuSTAR data were included in the DEM calculation for this feature (though including Hinode/XRT was found to not change the shown solution significantly). All three of the resulting DEMs for BP1 and BP2 are shown in the top row of Figure \ref{fig:bpdems}. They confirm the behaviour found in the NuSTAR spectral fits, with the increase in emission for BP1 between the two orbits, and confirm the presence of slightly hotter material in BP2, with a higher DEM for $\log_{10} T\gtrsim$ 6.2.
The bottom row of Figure \ref{fig:bpdems} plots the DEM for all 3 bright points: BP1 and BP2 during orbit 2 and BP3 during orbit 1. BP3 is fainter than BP1 and BP2, but its DEM has a similarly shaped fall for $\log_{10} T\gtrsim$ 6.2 to BP2,  for both of which the NuSTAR spectral fit found a similar temperature (which was higher than that found for BP1). All the bright point DEMs shown have peaks at $\log_{10} T\sim$ 5.7 and 6.15, a result which has also been found in previous DEM analyses of coronal bright points using EUV spectroscopy \citep{brosius,doschek}. These previous works and the DEMs presented in this paper all show no significant emission present above 4 MK. In our case, these DEMs benefit from using the X-ray data from NuSTAR and Hinode/XRT to constrain the higher temperature emission.

\par
\subsection{NuSTAR Non-thermal Upper Limits}
For strongest bright points, BP1 and BP2, the non-thermal upper limits were calculated when these were brightest (from orbit 2), again following the approach from Section \ref{upperlimits}. For BP1, an area of $35''$ square gives a volume of 1.65 $\times$ $10^{28}$ cm$^{3}$, a thermal energy of 2.71 $\times$ $10^{27}$ erg, and a heating power of 5.02 $\times$ $10^{24}$ erg s$^{-1}$ when the energy is divided by the NuSTAR observation time (540~s). The largest non-thermal upper limits for BP1 were about an order of magnitude smaller than this heating requirement. If there was some filling factor $<$ 1 and the area used here was an overestimate, then the heating requirement could be reduced but at best BP1 would still be at the very limit of being a possible non-thermally heated source. BP2, with an area of $15''$ square, gives a volume of 1.30 $\times$ $10^{27}$ cm$^{3}$, thermal energy of 3.71 $\times$ $10^{26}$ erg, and heating power of 1.52 $\times$ $10^{24}$ erg s$^{-1}$ when this energy is divided by the NuSTAR observation time (245~s). The upper limits on the non-thermal heating power are only slightly lower than this value for a very steep, almost mono-energetic, spectrum with a low energy cutoff of $\sim$ 3 keV. Again, the heating requirement could be shifted down by using a filling factor $<$ 1 and reducing the area. Therefore, it is possible that this BP2 could have been heated non-thermally, but this result is marginal.

\section{Jet}
\label{jet}
\begin{figure}    
   \centerline{\includegraphics[width=1.0\textwidth,clip=]{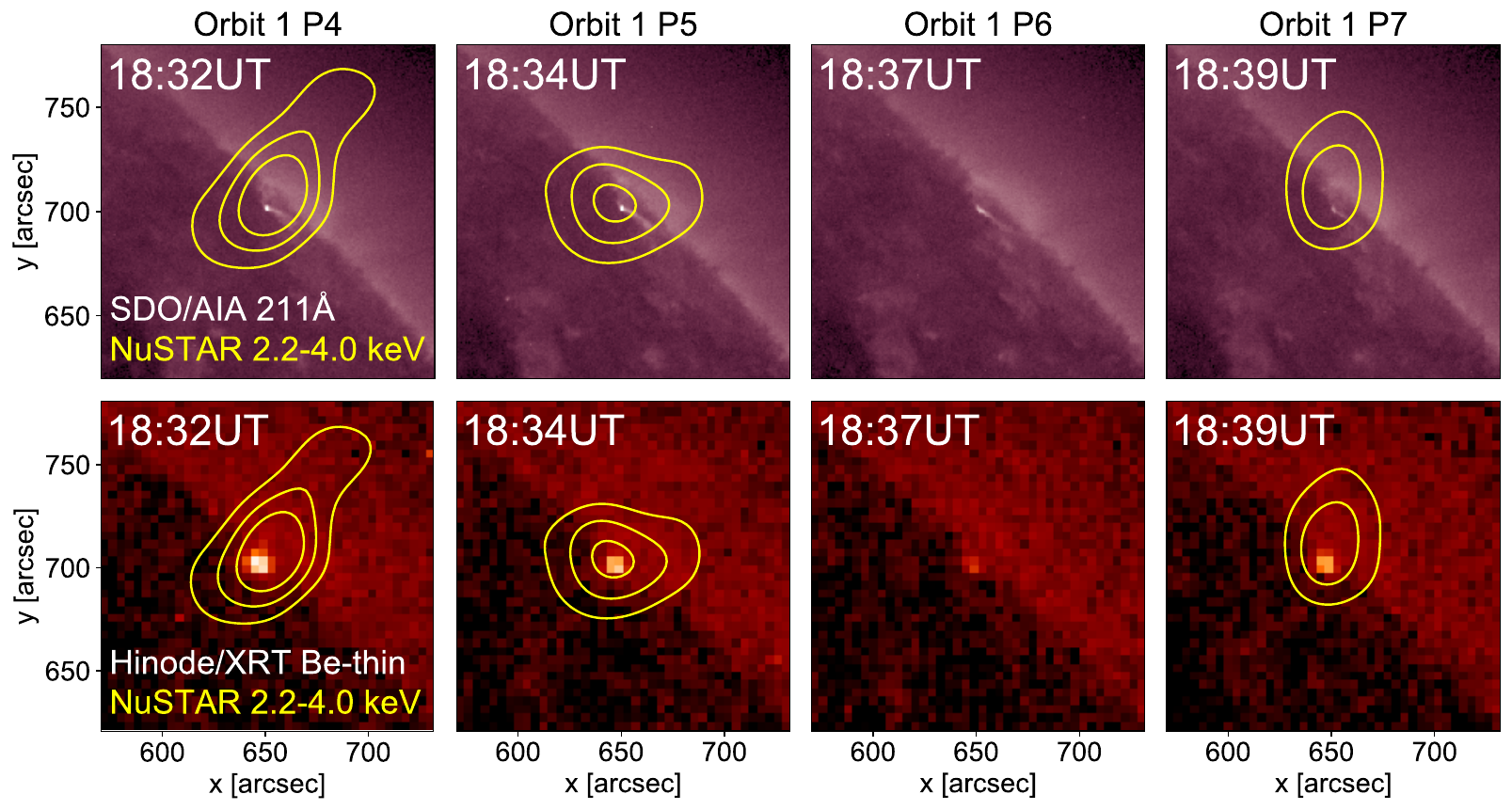}
              }
              \caption{SDO/AIA 211 Å (top) and Hinode/XRT Be-thin (bottom) images of the jet from the mid-times of the four NuSTAR pointings in the first orbit which captured it. Yellow contours represent NuSTAR FPMA + FPMB 2.2--4.0 keV, with the contours aligned with SDO/AIA and plotted at the same levels in all panels (5, 7, 9 $\times$ $10^{-4}$ counts s$^{-1}$).}
   \label{fig:trcontours}
   \end{figure}

A transient feature is observed at the top right of the orbit 1 NuSTAR mosaic (first panel of Figure \ref{fig:fullmaps}), caught in pointing P4 through 7 but gone by orbit 2. SDO/AIA images of this confirm that it is a compact jet which begins to brighten at around 18:27 UT, and has disappeared by 18:40 UT. SDO/AIA 211 Å and Hinode/XRT Be-thin images from the mid-times of each pointing are plotted in Figure \ref{fig:trcontours}, with aligned NuSTAR contours. From the SDO/AIA images, it can be seen that the configuration of this jet is atypical, with the jet material being ejected perpendicularly rather than radially outwards, implying that the overlying magnetic field is pushing it sideways. 

\begin{figure}    
   \centerline{\includegraphics[width=1.0\textwidth,clip=]{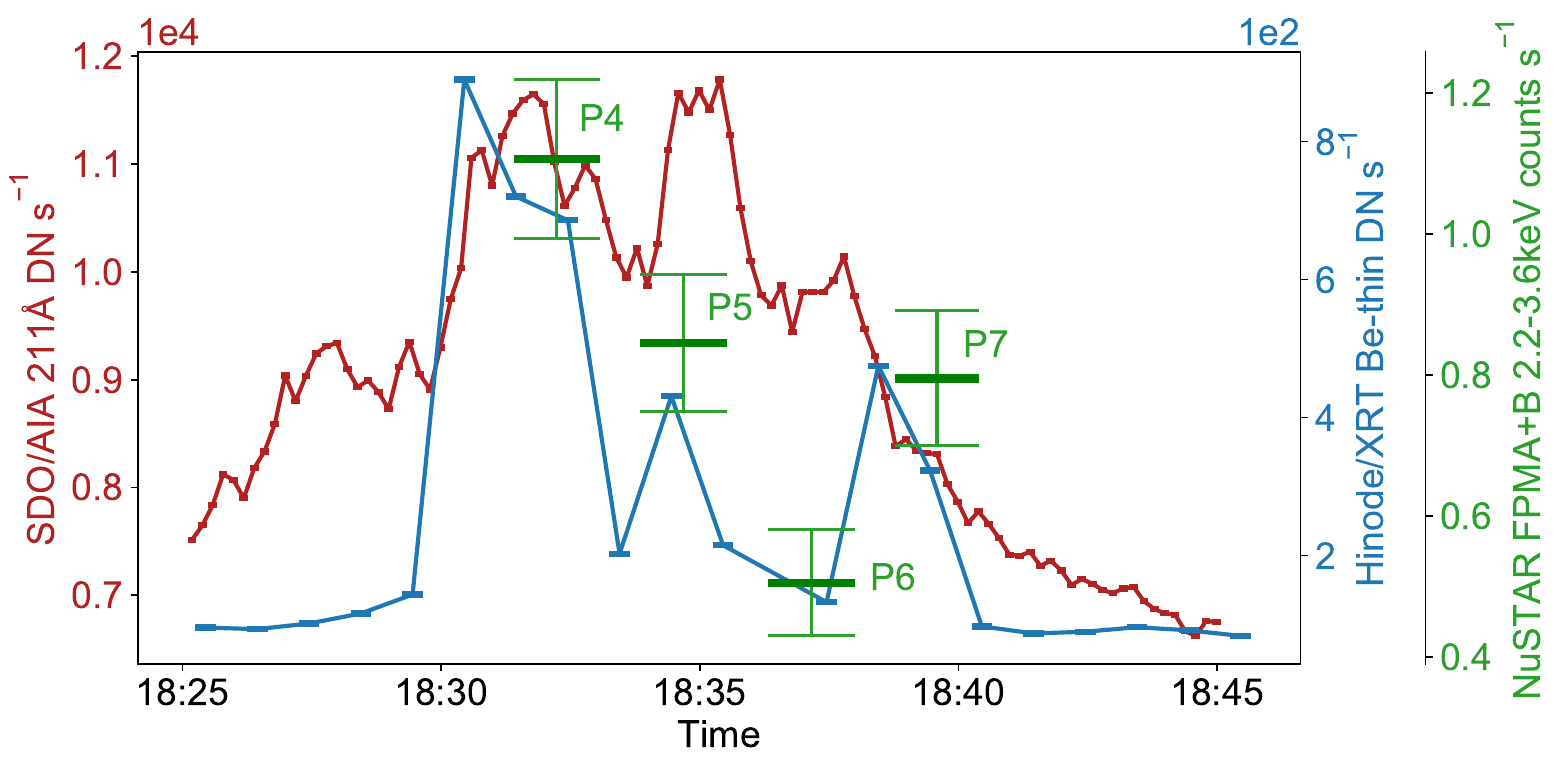}
              }
              \caption{SDO/AIA 211 Å and Hinode/XRT Be-thin lightcurves for the jet. The green lines indicate the NuSTAR FPMA + FPMB livetime-corrected count rates, and their corresponding errors, for the pointings that captured this feature.}
   \label{fig:trlightcurves}
   \end{figure}

The jet's lightcurves are shown in Figure \ref{fig:trlightcurves}, with the NuSTAR count rates for each pointing plotted for comparison. When NuSTAR sees the brightest emission from the jet (P4 in Figure \ref{fig:trlightcurves}) there is also a peak in both SDO/AIA 211 Å and Hinode/XRT Be-thin. Another peak in SDO/AIA 211Å coincides with P5, but this feature has decreased in brightness in both NuSTAR and Hinode/XRT. Again showing agreement with Hinode/XRT, though not SDO/AIA 211Å, the NuSTAR signal is at its lowest in P6 (making the feature almost indistinguishable from the background), before increasing in brightness in P7. This behaviour is also apparent in the Hinode/XRT images, where the feature appears brighter during the times of NuSTAR P4, 5, and 7 compared to P6. The jet is positioned far enough away from any chip gaps that this change in brightness in NuSTAR is genuine, as opposed to an effect of it moving in and out of detector gaps.

\begin{figure}    
   \centerline{\includegraphics[width=0.45\textwidth,clip=]{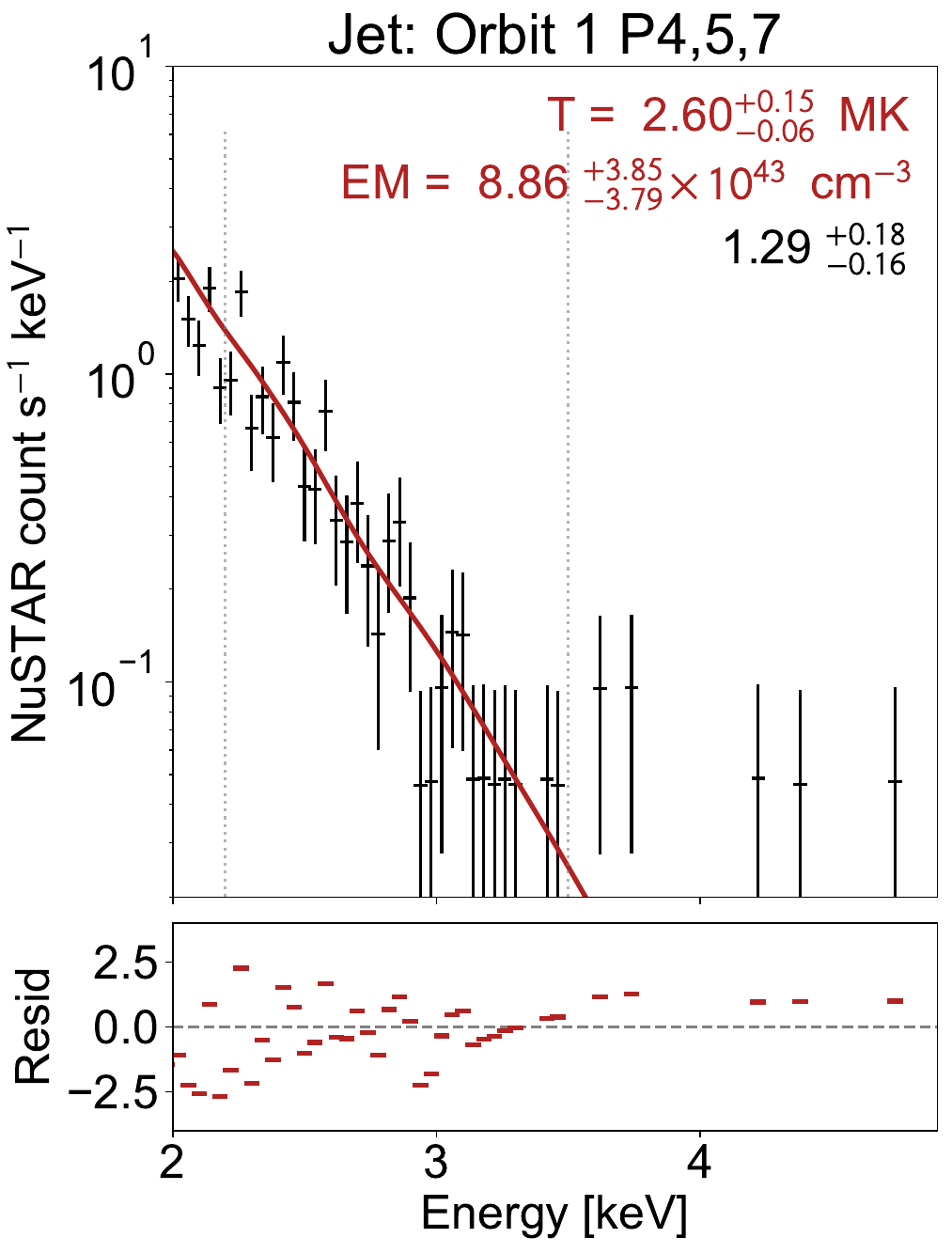}
              }
              \caption{NuSTAR spectral fit for the jet in orbit 1, simultaneously fitting FPMA and FPMB from P4 (18:31:24--18:33:04 UT), 5 (18:33:51--18:35:32 UT), and 7 (18:38:46--18:40:24 UT). Dotted lines indicate fitting range, and temperature and emission measure are marked on the plot. The number in black is the multiplicative constant introduced to account for systematic differences between FPMA and FPMB.}
\label{fig:toprightspec}
\end{figure}

\subsection{NuSTAR Spectral Analysis}
The NuSTAR spectra of P4, 5, and 7 were fit simultaneously and are shown in Figure \ref{fig:toprightspec}. P6 was not used due to the faintness of the NuSTAR emission. Though this source is clearly evolving in SDO/AIA 211Å, the NuSTAR spectra for each of the pointings were fit individually and it was found that there was no significant change in the temperature or emission measure throughout. The fit for the NuSTAR jet spectrum gives 2.60 MK and an emission measure of 8.86 $\times$ $10^{43}$ cm$^{-3}$. The isothermal model fits the spectrum well, with no indication of a higher temperature component or any non-thermal emission. This temperature lies in the range of the sensitivity of Hinode/XRT, but slightly above the peak in SDO/AIA 211Å, which might explain the different behaviour of SDO/AIA 211Å compared to Hinode/XRT and NuSTAR in Figure \ref{fig:trlightcurves}.

\subsection{NuSTAR Non-thermal Upper Limits}
\label{sec:jetlims}
\begin{figure}    
   \centerline{\includegraphics[width=0.8\textwidth,clip=]{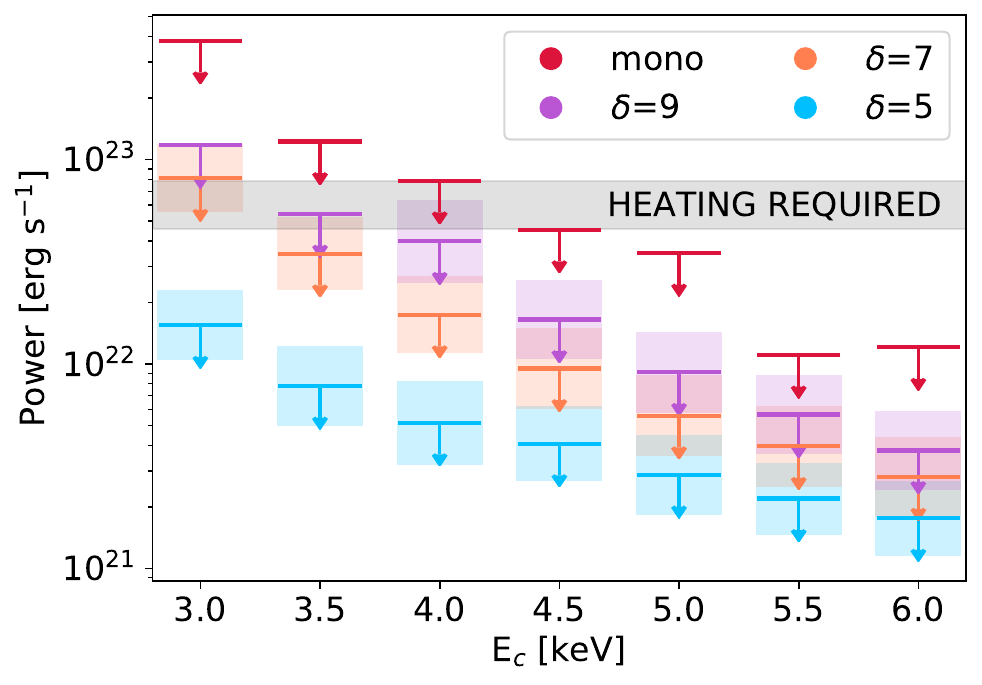}
              }
              \caption{Upper limits on the non-thermal heating power for a range of $E_{c}$ and $\delta$ (= 5, 7, 9, and a mono-energetic beam) values. The shaded regions indicate the $\pm$ 1 $\sigma$ range of the Gaussian distribution fitted to the upper limit results. The grey shaded area indicates the heating requirement dictated by the NuSTAR thermal emission, with the range determined from the uncertainties on the fit temperature and emission measure.}
   \label{fig:upperlims}
   \end{figure}

From an SDO/AIA 211Å image of the jet, we get an area of $3''$ square and a volume of 1.04 $\times$ $10^{25}$ cm$^{3}$. Therefore, taking the temperature and emission measure values from the NuSTAR spectral fits, the thermal energy of the jet is 3.26 $\times$ $10^{25}$ erg. 
Upper limits on the possible non-thermal emission that could be present and remain undetected by NuSTAR were calculated for this event using the method discussed in Section \ref{upperlimits}, with the results plotted in Figure \ref{fig:upperlims}. Note that, for the case of the mono-energetic beam and for $\delta$ = 9 and $E_{c}$ $<$ 4 keV the results were more well-defined that for the other cases, and therefore no spread is indicated in Figure \ref{fig:upperlims}. The heating power required for the jet over the time range considered here (obtained by dividing the thermal energy by the NuSTAR observation time of 528~s) is 6.18 $\times$ $10^{22}$ erg s$^{-1}$, which is marked on Figure \ref{fig:upperlims} in comparison to the non-thermal upper limits. The non-thermal power would have to be greater than or equal to the heating requirement in order for the feature to have been heated through by accelerated electrons. Therefore, it can be concluded that only if the non-thermal emission was very steep, almost mono-energetic, between 3 and 4 keV could it power the required heating as determined from the NuSTAR thermal emission.

\section{Bright Limb Emission}
\label{limb}
The brightest, and also largest, source that appears in the NuSTAR mosaics in Figure \ref{fig:fullmaps} is the emission from the east limb. From the SDO/AIA and Hinode/XRT images, this feature is two different sources: a bright loop, and surrounding diffuse emission. Looking back a solar rotation before this observation, this area was the site of an active region that, though too faint to be given a NOAA identification number, was detected with the Spatial Possibilities Clustering Algorithm, as SPoCA 22053 \citep{delouille}. Therefore, this bright emission is likely due to the presence of a decayed active region, and so we label it DAR loop.

This source is captured fully by P11 and partially by P10 and 20 in both NuSTAR orbits. This limb emission is relatively bright and extended, meaning that the noise is not as big of an issue as it is with the other features presented in this paper. However, there are other factors which complicate the fitting of its NuSTAR spectra. Firstly, in each of the pointings this source is extended enough to be positioned over multiple NuSTAR detector quadrants, which have varying responses. Also, as mentioned previously, the NuSTAR emission is a combination of the bright loop and the surrounding diffuse emission. Therefore, in order to investigate the properties of the emission that originates only from the bright loop in SDO/AIA and Hinode/XRT, the brightest section of the NuSTAR emission was chosen for the fitting. However, this bright loop is not ideally positioned in most of the pointings that capture this area, lying over the detector gap or just off the edge of the detector in several of these pointings. The bright region was best observed in Orbit 1 P10 and 11 with FPMA, and therefore only these pointings were combined for the spectral fitting. The NuSTAR spectrum was again fit with an isothermal model, giving a temperature of 2.53 MK and an emission measure of 9.62 $\times$ $10^{44}$ cm$^{-3}$. Because it is only possible to do a NuSTAR spectral fit for one point in time for this feature, its HXR evolution cannot be compared to its EUV and SXR evolution in SDO/AIA and Hinode/XRT.

\section{Comparison of Thermal Properties}\label{thermcomp}

\begin{table}[ht]
\caption{A summary of the isothermal models fit to the NuSTAR spectra for all of the features. Results from different times throughout the NuSTAR observation are given, in the appropriate cases.}
\label{specresults}
\begin{tabular}{cccc}     
  \hline                   
Feature & Orbit/Pointing & Temperature & Emission Measure \\
  &  &  MK & $\times$ $10^{43}$ cm$^{-3}$\\
  \hline

EFR (Section \ref{efr}) & Orbit 1 P12,13 & $2.54${\raisebox{0.5ex}{\tiny$^{+0.04}_{-0.03}$}} & $34.2${\raisebox{0.5ex}{\tiny$^{+8.6}_{-8.5}$}} \\
& Orbit 1 P18,19 & $2.54${\raisebox{0.5ex}{\tiny$^{+0.04}_{-0.03}$}} &$19.3${\raisebox{0.5ex}{\tiny$^{+5.1}_{-4.9}$}} \\
& Orbit 2 P12,13 & $2.57${\raisebox{0.5ex}{\tiny$^{+0.04}_{-0.03}$}} &$61.0${\raisebox{0.5ex}{\tiny$^{+12.3}_{-12.4}$}} \\
& Orbit 2 P18,19 & $2.63${\raisebox{0.5ex}{\tiny$^{+0.10}_{-0.05}$}} &$24.0${\raisebox{0.5ex}{\tiny$^{+7.2}_{-7.2}$}} \\
BP1 (Section \ref{bps}) & Orbit 1 P14,15,16,17 & $2.55${\raisebox{0.5ex}{\tiny$^{+0.05}_{-0.03}$}} & $11.2${\raisebox{0.5ex}{\tiny$^{+3.0}_{-3.0}$}}\\
& Orbit 2 P14,15,16,17 & $2.53${\raisebox{0.5ex}{\tiny$^{+0.02}_{-0.01}$}}  &$40.7${\raisebox{0.5ex}{\tiny$^{+5.4}_{-5.4}$}}  \\
BP2 (Section \ref{bps}) & Orbit 2 P16,17 & $3.22${\raisebox{0.5ex}{\tiny$^{+0.06}_{-0.04}$}} & $5.98${\raisebox{0.5ex}{\tiny$^{+1.14}_{-1.14}$}}  \\
BP3 (Section \ref{bps}) & Orbit 1 P 13,14,17,18 & $3.22${\raisebox{0.5ex}{\tiny$^{+0.10}_{-0.16}$}} & $1.33${\raisebox{0.5ex}{\tiny$^{+0.38}_{-0.37}$}} \\
& Orbit 2 P 13,14,17,18 & $2.56${\raisebox{0.5ex}{\tiny$^{+0.09}_{-0.04}$}} & $5.10${\raisebox{0.5ex}{\tiny$^{+1.86}_{-1.88}$}}  \\
QS Loops (Section \ref{bps}) & Orbit 1 P13,14,17,18 & $2.51${\raisebox{0.5ex}{\tiny$^{+0.03}_{-0.38}$}} & $10.3${\raisebox{0.5ex}{\tiny$^{+20.8}_{-2.3}$}}  \\
& Orbit 2 P13,14,17,18 & $2.07${\raisebox{0.5ex}{\tiny$^{+0.12}_{-0.04}$}} &  $63.9${\raisebox{0.5ex}{\tiny$^{+31.2}_{-32.0}$}}  \\
Jet (Section \ref{jet}) & Orbit 1 P 4,5,7 & $2.60${\raisebox{0.5ex}{\tiny$^{+0.15}_{-0.06}$}} & $8.86${\raisebox{0.5ex}{\tiny$^{+3.85}_{-3.79}$}}  \\
DAR Loop (Section \ref{limb}) & Orbit 1 P10,11 & $2.53${\raisebox{0.5ex}{\tiny$^{+0.04}_{-0.14}$}} & $96.2${\raisebox{0.5ex}{\tiny$^{+49.6}_{-21.4}$}}  \\
  \hline
\end{tabular}
\end{table}

\begin{figure}    
   \centering
   \includegraphics[width=1.0\textwidth,clip=]{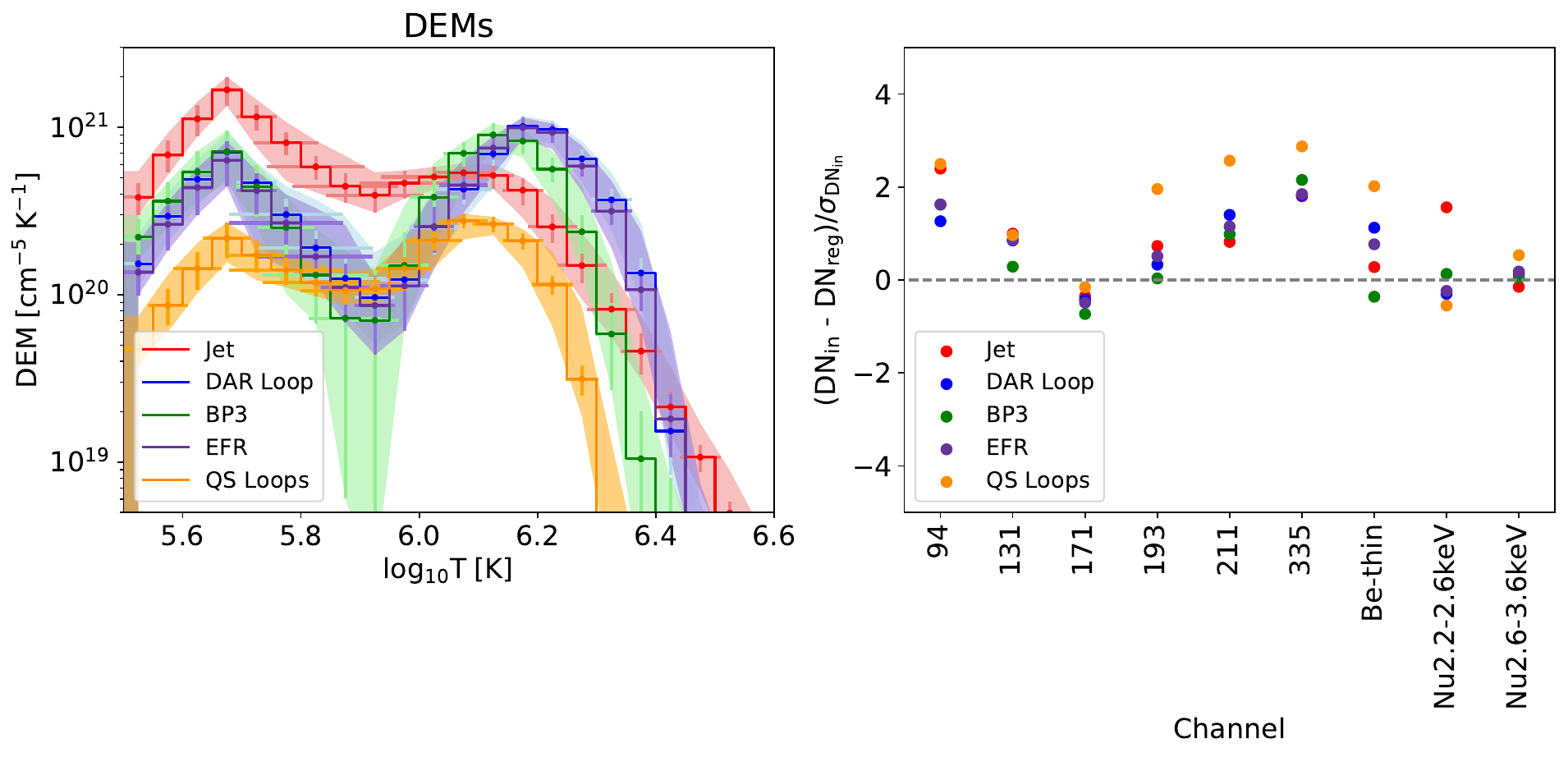}
              \caption{(Left) DEMs for some of the quiet Sun features studied in this paper, with the shaded region indicating the uncertainty in the DEM solutions comparison plot. (Right) The corresponding normalised residuals in data space for each feature's recovered DEM.}
   \label{fig:demcomp}
   \end{figure}

A variety of quiet Sun features have been analysed in this survey, with their X-ray emission detected for the first time by a focusing optics imaging spectrometer, allowing their X-ray spectra to be fitted. The results of fitting an isothermal model to all these features' NuSTAR spectra are given in Table \ref{specresults}. The temperatures found from the NuSTAR spectral fits for all of the features lie in a narrow temperature range between 2.0--3.2 MK. This is due to these events being at the limit of NuSTAR's temperature sensitivity, and there is so little hotter material that the spectra are dominated by these cooler sources, appearing effectively isothermal. This is consistent with previous studies that used EUV and SXR data, which found that quiet Sun features like bright point are generally $<$ 3 MK \citep{doschek,alexander,kariyappa}. BP2 and BP3 are slightly hotter than have been previously studied, with the NuSTAR spectral fits giving 3.2 MK. Although NuSTAR is more sensitive to higher temperature emission, there is so little of it in these features that the spectra are dominated by these cooler sources, appearing effectively isothermal, capturing the 2--3 MK peak of the DEM.
\par
The DEMs for several of the features are plotted together in Figure \ref{fig:demcomp} and all show very similar two peak structures, with peaks around $\log_{10} T\sim $ 5.7 and 6.1--6.2. This lower peak has been seen before in DEM analysis using EUV spectroscopy \citep{doschek,brosius} and in our observations is dominated by the emission seen in SDO/AIA 131\AA. Increasing the lower temperature limit our DEMs are calculated over (towards $\log_{10} T\sim $ 5.9) has minimal change on the higher temperature DEM component $\log_{10} T\gtrsim $6.1 and only produces a large discrepancy to the observed SDO/AIA 131\AA. Removing this channel from our DEM calculation again has minimal change to the higher temperature peak and tail of the DEMs. The higher temperature peak is slightly lower for the QS loops, jet and BP3, about $\log_{10} T\sim $ 6.1 compared to the EFR and DAR loops. Both the EFR and DAR loop have very similar DEMs, which may be just coincidence, but is curious given that one is the very start of an active region and the other the decayed remains. All DEMs fall off rapidly above this peak, highlighting that very little material has been heated to higher temperatures. The jet DEM is slightly flatter than the others, indicating possibly more hotter material than the other features but this was not confirmed by the NuSTAR spectral fit. However this was a small, faint and short duration event so would have been hampered by NuSTAR's limited detector throughput, which would not have been helped by a longer dwell observation.

Difficulties also arise in this analysis when working with SDO/AIA data because none of the channels have a peak in sensitivity in the 2--3MK range. As a result, the SDO/AIA 211 Å lightcurves for these sometimes do not show behaviour consistent with Hinode/XRT or NuSTAR. Previous analysis of microflares observed with NuSTAR (for example, \cite{cooper1,cooper2}) has used the SDO/AIA Fe XVIII proxy channel \citep{dz}. Unfortunately, the temperatures of these quiet Sun features are too low for this to be useful. However, Hinode/XRT has sensitivity in a similar temperature range to NuSTAR, and makes a useful comparison.

\section{Discussion and Conclusions}
\label{conclusions}
In this paper, we have presented the first survey of quiet Sun features in HXRs observed during solar minimum. NuSTAR's full-disk solar mosaic mode allowed for a range of different types of features to be observed. In these two mosaics, NuSTAR observed steady features, such as bright points and an EFR, but also captured a transient jet. This is the first observation of these types of features using a HXR focusing telescope. The mosaics also reveal large-scale sources (the diffuse sources in Figure \ref{fig:fullmaps}), diagnostically important for investigating the heating of the diffuse corona.
\par
As summarised in Section \ref{thermcomp} and Table \ref{specresults}, we find the features' temperatures lie in the range 2.0--3.2 MK, capturing the sharp fall off in their DEMs. We find no evidence of a higher temperature or non-thermal component present in their X-ray spectra. We have used EUV and SXR data from SDO/AIA and Hinode/XRT in addition to NuSTAR to investigate the temperature evolution of the quiet Sun features, including successfully reconstructing DEMs which combine data from all three of these instruments. The DEM solutions for these quiet Sun features show no evidence of emission above 4 MK, a result achieved by using X-ray data in the DEM calculation to constrain the solution at high temperatures.
\par
As all of the NuSTAR spectra were adequately fitted with an isothermal model, only non-thermal upper limits were found for some of the features. In most cases, it was found that the possible non-thermal component was not sufficient to produce the required heating. The feature that was the best candidate for non-thermal heating was the jet. However, even this would require a very steep (effectively mono-energetic) non-thermal distribution with a low energy cutoff between 3--4 keV.
\par
From the spectral, DEM, and non-thermal upper limit analysis performed here, it can be concluded that no higher temperature or non-thermal sources were found in this quiet Sun data. However, if any such components were present they  would be very faint, and therefore NuSTAR does not have the sensitivity required to detect them in the short 100 s mosaic pointings combined with its limited throughput. Higher temperature or non-thermal components would only be detectable if they were relatively strong, or in longer duration observations of non-transient features.
\par
The work presented in this paper used the first NuSTAR quiet Sun campaign from the recent solar minimum. Additional data sets were taken throughout the solar minimum (2018--2020) in both the full-disk mosaic mode as well as longer dwells, in which pointing was not changed. In these dwells, any bright points would be observed for several hours over multiple orbits. These longer observing campaigns could increase the chances of detecting more energetic HXR emission from the quiet Sun, and of capturing more atypical harder sources. Having observations of quiet Sun features over a longer period of time will also mean that a more rigorous investigation of their temporal evolution in HXRs will be possible. The NuSTAR quiet Sun dwell data will be used to further the work presented here, and will be the subject of future papers. However, shorter time-scale variability in the HXR emission from quiet Sun features such as these may remain difficult to detect until there is a dedicated solar X-ray instrument with higher sensitivity and throughput.

\begin{acks}

This paper made use of data from the NuSTAR mission, a project led by the California Institute of Technology, managed by the Jet Propulsion Laboratory, funded by the National Aeronautics and Space Administration. We thank the NuSTAR Operations, Software and Calibration teams for support with the execution and analysis of these observations. This research made use of the NuSTAR Data Analysis Software (NUSTARDAS) jointly developed by the ASI Science Data Center (ASDC, Italy), and the California Institute of Technology (USA). Hinode is a Japanese mission developed and launched by ISAS/JAXA, with NAOJ as domestic partner and NASA and UKSA as international partners. It is operated by these agencies in co-operation with ESA and NSC (Norway).  AIA on the Solar Dynamics Observatory is part of NASA’s Living with a Star program. This research has made use of SunPy, an open-source and free community-developed solar data analysis package written in Python \citep{2015CS&D....8a4009S}.  This research made use of Astropy, a community-developed core Python package for Astronomy \citep{2018AJ....156..123A, 2013A&A...558A..33A}.  The authors would also like to thank Kathy Reeves for co-ordinating the Hinode/XRT observations.
\end{acks}

\begin{fundinginformation}
SP acknowledges support from the UK's Science and Technology Facilities Council (STFC) doctoral training grant (ST/T506102/1). IGH acknowledges support from a Royal Society University Fellowship (URF/R/180010) and STFC grant (ST/T000422/1).
\end{fundinginformation}

\begin{dataavailability}
All the data used in this paper are publicly available. In particular, NuSTAR via the  \href{https://heasarc.gsfc.nasa.gov/db-perl/W3Browse/w3table.pl?tablehead=name=numaster&Action=More+Options}{NuSTAR Master Catalog} with the OBSIDs 90410101001 - 90410125001 and 90410201001 - 90410225001.
\end{dataavailability}

\begin{ethics}
\begin{conflict}
The authors declare that they have no conflicts of interest.
\end{conflict}
\end{ethics}

\end{article} 

\end{document}